\documentclass[twocolumn]{aastex701}

\usepackage{hyperref}

\usepackage{natbib}
\usepackage{multirow}
\usepackage{color}
\usepackage{bm}
\usepackage[normalem]{ulem}
\usepackage{amsmath}
\usepackage{url}

\newcommand{\eg}{e.g., }
\newcommand{\ie}{i.e., }

\newcommand{\Msun}{M_{\sun}}

\newcommand{\Mej}{M_{\rm ej}}
\newcommand{\vej}{v_{\rm ej}}
\newcommand{\Md}{M_{\rm d}}

\newcommand{\kdi}{k^{\rm assoc}_{\rm di}}

\def\ion#1#2{{\rm #1}~{\sc #2}}

\shorttitle{$r$-process dust in kilonovae}
\shortauthors{N. Domoto et al.}

\begin{document}

\title{Heavy-element dust as an explanation for the late-time kilonova spectra}

\correspondingauthor{Nanae Domoto}

\author[orcid=0000-0002-7415-7954]{Nanae Domoto}
\affiliation{Research Center for the Early Universe, Graduate School of Science, The University of Tokyo, 7-3-1 Hongo, Bunkyo-ku, Tokyo 113-0033, Japan}
\email[show]{ndomoto@g.ecc.u-tokyo.ac.jp}

\author[orcid=0000-0002-2502-3730]{Kenta Hotokezaka}
\affiliation{Research Center for the Early Universe, Graduate School of Science, The University of Tokyo, 7-3-1 Hongo, Bunkyo-ku, Tokyo 113-0033, Japan}
\affiliation{Max Planck Institute for Gravitational Physics (Albert Einstein Institute), Am M\"uhlenberg 1, Potsdam-Golm 14476, Germany}
\email{kentah@g.ecc.u-tokyo.ac.jp}

\author[orcid=0000-0002-5981-1022]{Daniel Kasen}
\affiliation{Department of Astronomy, University of California, 501 Campbell Hall, Berkeley, CA 94720, USA}
\affiliation{Department of Physics, University of California, 366 Physics North MC 7300, Berkeley, CA 94720, USA}
\affiliation{Nuclear Science Division, Lawrence Berkeley National Laboratory, 1 Cyclotron Road, Berkeley, CA 94720, USA}
\email{kasen@berkeley.edu}



\begin{abstract}
Neutron star mergers are a leading site of $r$-process, producing radioactively powered optical and infrared transients known as kilonovae. 
Observations of the kilonovae AT2017gfo, associated with the gravitational-wave event GW170817, and AT2023vfi, associated with GRB 230307A, have enabled measurements of the mass of ejected $r$-process material and identification of heavy elements in the ejecta.
However, late-time observations reveal strong infrared emission with temperature below 1000 K, which is difficult to explain by atomic absorption and emission processes alone. 
In this paper, we propose that kilonova ejecta provide conditions favorable for the formation of dust grains composed of refractory $r$-process elements including Zr, W, and Os.
Assuming a simplified, empirically derived single power-law temperature evolution and using W reaction rate coefficients as a proxy, our kinetic dust formation calculations indicate that dust could form efficiently, particularly in slow ejecta. This stands in contrast to a previous study that relied on a classical nucleation framework. By performing radiative transfer simulations that incorporate dust formation, we demonstrate that, if formed, $r$-process dust could naturally explain the observed late-time infrared emission, although refining ejecta parameters and geometry will be necessary to match specific observational cases. The formation and abundance of $r$-process dust are highly sensitive to ejecta mass, composition, and expansion velocity. Infrared emission from $r$-process dust may therefore potentially serve a new probe of heavy-element production in neutron star mergers.
\end{abstract}


\keywords{\uat{R-process}{1324} -- \uat{Neutron stars}{1108} -- \uat{Dust physics}{2229}}

\section{Introduction}
\label{sec:intro}
Binary neutron star mergers are considered promising sites of rapid neutron capture ($r$-process) nucleosynthesis \citep[e.g.,][]{LS1974, Eichler1989, Meyer1989, Freiburghaus1999, Goriely2011a, Korobkin2012, Wanajo2014}. Radioactive decay of freshly synthesized nuclei in the ejected neutron-rich material powers electromagnetic emission, called a kilonova \citep{LiPaczynski1998, Metzger2010, Roberts2011}.
In 2017, associated with the detection of gravitational waves (GWs) from a neutron star merger \citep[GW170817;][]{Abbott2017a}, an electromagnetic counterpart was identified \citep[AT2017gfo;][]{Abbott2017b}.
The observed properties of AT2017gfo in ultraviolet (UV), optical, and near-infrared (NIR) wavelengths are consistent with the theoretical expectations of a kilonova \citep[e.g.,][]{Arcavi2017, Coulter2017, Evans2017, Kasliwal2017Sci, Pian2017, Smartt2017, Utsumi2017, Valenti2017}.
This electromagnetic counterpart has provided evidence that neutron star mergers are sites of $r$-process nucleosynthesis \citep[e.g.,][]{Kasen2017, Perego2017, Shibata2017, Tanaka2017, Kawaguchi2018, Rosswog2018}, enabling measurements of the mass of ejected $r$-process material and the identification of heavy elements in the ejecta \citep{Kasen2017,Watson2019, Domoto2021, Domoto2022, Sneppen2023, Hotokezaka2023,Mulholland2026MNRAS}.

Recently, another kilonova candidate AT2023vfi was discovered in association with a gamma-ray burst GRB~230307A \citep{Levan2024, Yang2024}.
JWST observations of AT2023vfi at 29\,days revealed a strong infrared continuum spectrum peaking at $\sim 5\,{\rm \mu m}$, which is well described by blackbody radiation with a temperature of $\approx 660$\,K \citep{Gillanders2025}. 
In fact, similar late-time red emission was also observed in AT2017gfo several tens of days after the merger \citep{Villar2018, Kasliwal2022}.
At such low temperatures, atomic opacity from bound-bound transitions of $r$-process elements is expected to be insufficient to produce the observed emission \citep{Kasen2013ApJ, Fontes2020, Tanaka2020, Flors2026, Jerkstrand2025, Pognan2026}.

Thermal emission from newly formed dust grains would naturally produce such an infrared continuum \citep[see, e.g.,][]{Gall2017ApJ,Arunachalam2025arXiv}. 
Previous studies on dust formation in kilonovae, however, have concluded that the formation of $r$-process grains is unlikely because the nucleation rate is much lower than the expansion rate based on kinetic nucleation theory \citep{Takami2014}. 
On the other hand, it has been discussed that the classical picture of nucleation theory is not applicable to dust formation where the formation of small clusters determines the nucleation rate \citep[e.g.][]{Draine1979Ap&SS,Jun2022ARPC}. Therefore, it is crucial to solve a molecular reaction network for cluster formation in an expanding medium to accurately capture these microphysical processes \citep{Cherchneff2010,Goumans2012MNRAS,Lazzati2016ApJ,Sluder2018}.

In this paper, we study the formation and emission of dust composed of $r$-process refractory elements in kilonova ejecta.
In Section~\ref{sec:mass}, we first give an estimate for the dust mass required to explain the late-time emission of AT2023vfi and discuss the expected abundance of refractory elements in kilonova ejecta. 
In Section~\ref{sec:mot}, we show the condensation temperatures of these elements, demonstrating that kilonova ejecta provide favorable conditions for dust formation.
Then, in Section~\ref{sec:form}, we investigate the dust formation by calculating the kinetic evolution of $r$-process dust grains.
In Section~\ref{sec:rad}, we perform radiative transfer simulations of kilonovae incorporating dust formation model, and show $r$-process dust can naturally explain the late-time kilonova spectrum.
Finally, we discuss the broader implications of our results for dust formation, $r$-process, and kilonova observations in Section~\ref{sec:discussion}, and give our conclusions in Section~\ref{sec:conclusion}.

\begin{figure*}
    \centering
    \includegraphics[width=\linewidth]{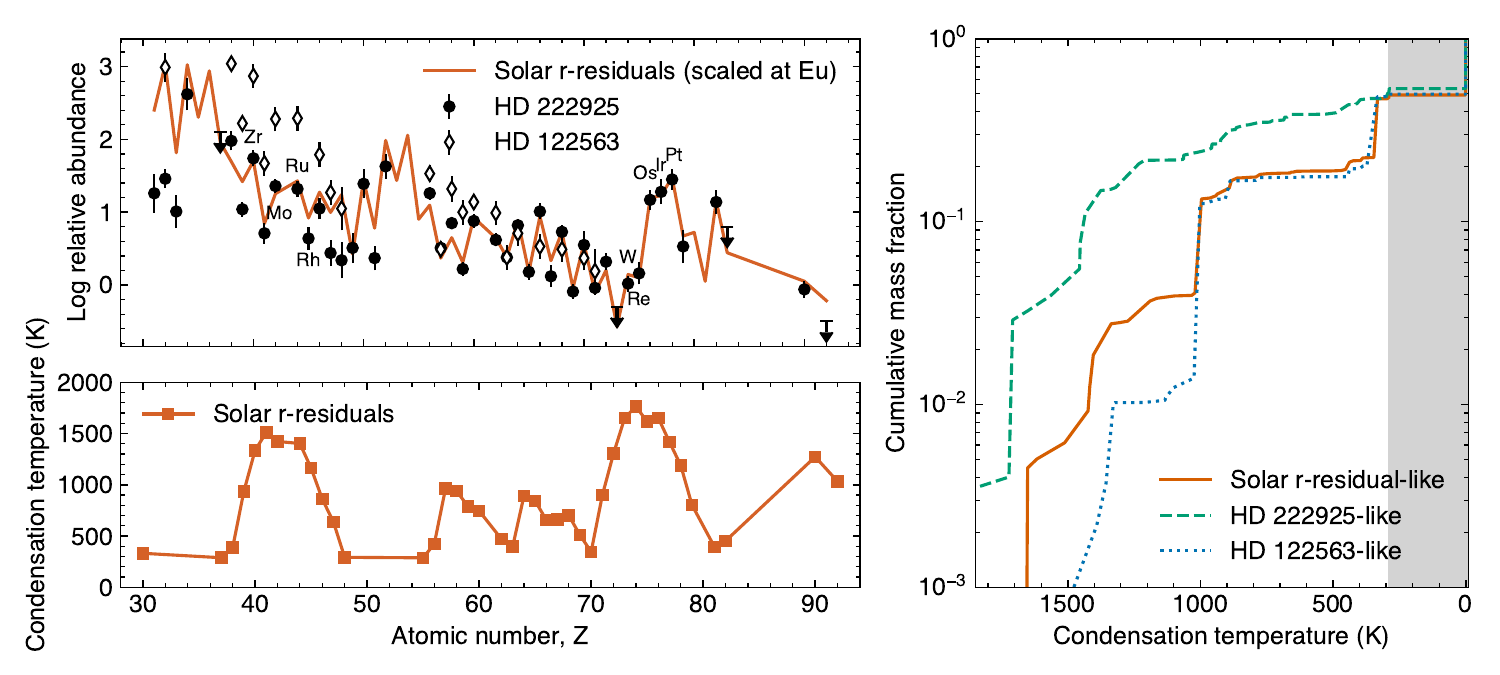}
    \caption{
    Top left: relative abundances of the solar $r$-residuals \citep{Prantzos2020} and of metal-poor $r$-process-enhanced stars HD~222925 \citep{Roederer2022} and HD~122563 \citep{Cowan2005, Honda2006, Roederer2012}. The relative abundances are scaled at Eu ($Z=63$).
    Bottom left: condensation temperatures of heavy elements assuming the solar $r$-residual pattern under kilonova conditions at 10\,days (see also Figure~\ref{fig:cond} for W).
    Right: cumulative mass fractions as a function of condensation temperature for different abundance models. The models are constructed to reproduce abundance patterns similar to the solar $r$-residuals or those of metal-poor $r$-process-enhanced stars (Figure~\ref{fig:elem-model} in Appendix \ref{sec:abun}). The gray shading indicates the fraction of elements for which condensation temperatures cannot be calculated owing to the lack of relevant data \citep{Alcock1984}. These elements are expected to lie at the low-temperature end \citep{Lodders2003, Wood2019}.
    }
    \label{fig:element}
\end{figure*}

\section{Dust mass estimate for AT2023vfi}\label{sec:mass}
Assuming that dust grains with a total mass of $\Md$ form in kilonova ejecta, the optical depth of the grains, $\tau_{\rm d}$, can be estimated as
\begin{equation}
	\tau_{\rm d} \approx n_{\rm d}\sigma_{\rm d}R_{\rm ej}
			   = \frac{9}{16\pi}\frac{\Md}{\rho_{\rm d} a v_{\rm ej}^2t^2} Q_{\rm abs}(\lambda),
	\label{eq:tau}
\end{equation}
where $R_{\rm ej}=\vej t$ is the ejecta radius, and $n_{\rm d}=3\Md/(4\pi R_{\rm ej}^3m_{\rm d})$ and $\sigma_{\rm d, abs}=Q_{\rm abs}(\lambda)\pi a^2$ are the number density and absorptive cross section of dust grains, respectively.
The mass of a single grain is $m_{\rm d}=4\pi a^3\rho_{\rm d}/3$, assuming a sphere of grain radius $a$ and uniform density $\rho_{\rm d}$.
The wavelength-dependent absorption efficiency of grains is approximated as $Q_{\rm abs}(\lambda) \approx a/\lambda$ for $a<\lambda$ otherwise 1.
For small-size grains with $a\ll \lambda=\mathcal{O}(1\,{\rm \mu m})$, the dust optical depth is independent of grain size. As will shown by equation~(\ref{eq:a_c}) in Section \ref{sec:res}, the typical grain size that may form in kilonova ejecta is expected to be the order of nanometers. Thus, at $\lambda=3\,\mu$m, the optical depth is given by
\begin{equation}
	\begin{aligned}
	\tau_{\rm d} &\approx 1 \\
		&\bigg(\frac{\Md}{10^{-3}\,\Msun}\bigg)
		\bigg(\frac{\rho_{\rm d}}{20\,{\rm g\,cm^{-3}}}\bigg)^{-1} 
		\bigg(\frac{\vej}{0.1\,c}\bigg)^{-2}
		\bigg(\frac{t}{29\,{\rm day}}\bigg)^{-2}
	\label{eq:tau1}
	\end{aligned}
\end{equation}
This indicates that a dust mass of $>6\times 10^{-4}\,\Msun$ is sufficient to produce an infrared continuum with $\vej=0.08\,c$ \citep{Gillanders2025}.
While previous studies have considered carbonaceous grains, such grains are unlikely to form in kilonova ejecta due to the low abundance of light elements \citep{Gall2017ApJ, Arunachalam2025arXiv}. 

Refractory elements including Zr, Mo, W, Re, Os, and Ir are expected to be abundant in kilonova ejecta. These elements have high condensation temperatures of $T\approx1800$\,K, below which they can condense into metallic grains, as detailed in the following section.
The abundance patterns of the solar system and metal-poor stars indicate that refractory elements comprise a few to 10\% of the $r$-process as shown in Figure~\ref{fig:element}.
The ejecta mass in neutron star mergers is inferred to be of order $5\times10^{-2}\,\Msun$, such that only a modest fraction of the refractory elements needs to condense into dust grains to account for the late-time infrared emission.

We note that the simplified $Q_{\rm abs}(\lambda)\approx a/\lambda$ relation represents a baseline approximation. While pure bulk metals typically exhibit high infrared reflectivity that steepens the wavelength dependence of the absorption efficiency, refractory-element dust in rapidly cooling and expanding ejecta is highly unlikely to condense into chemically pure crystalline grains. Instead, it will consist of complex mixtures of various co-condensing metals, expected to form non-crystalline solid structures, analogous to the predominantly amorphous silicate dust observed in circumstellar and interstellar environments \citep[e.g.,][]{Kemper2004}. The lack of crystallinity strongly suppresses free-electron reflectivity. Given the absence of experimental optical constants for non-crystalline $r$-process alloys in the literature, the standard $Q_{\rm abs}(\lambda) \approx a/\lambda$ approximation serves as a practical baseline for this study.

\begin{figure}
    \centering
    \includegraphics[width=\linewidth]{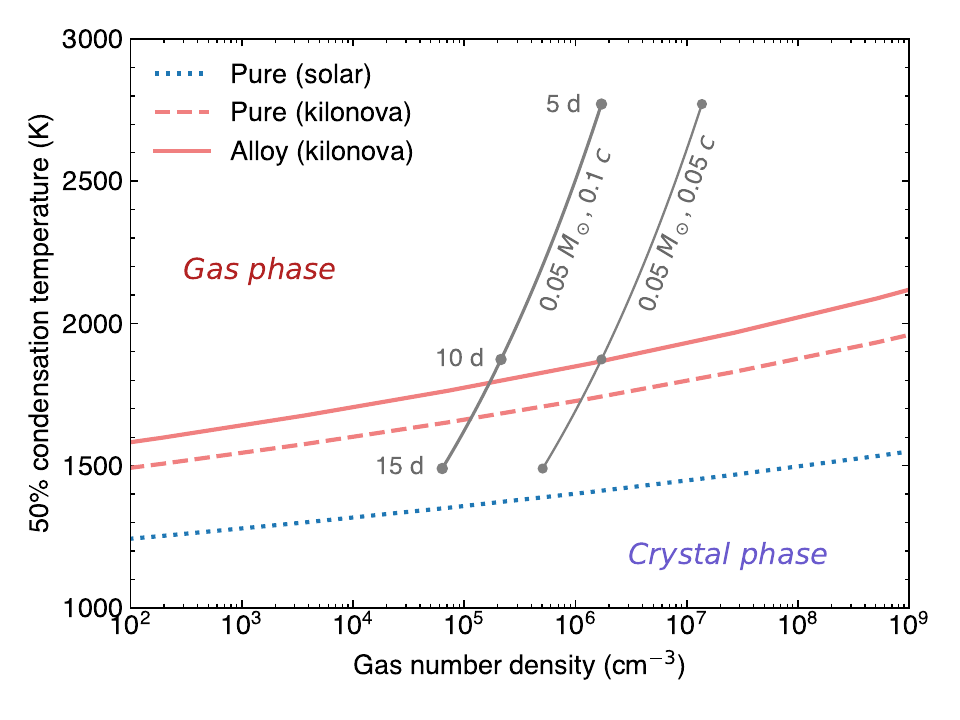}
    \caption{
    50\% condensation temperatures of W as a function of gas number density.
    The blue-dotted and red-dashed curves show pure W assuming the solar abundance \citep{Lodders2009} and the solar $r$-residuals \citep{Prantzos2020} in kilonova ejecta, respectively, while the red-solid curve shows an alloy of the third $r$-process peak elements with the solar $r$-residuals in kilonova ejecta.
    The gray curves indicate the evolutionary tracks of the ejecta with $\Mej=0.05\,\Msun$ and $\vej=0.1\,c$ (thick) and $0.05\,c$ (thin), assuming a condensable-element mass fraction of $\approx3$\% (see Section \ref{sec:kn_model}).
    The ejecta temperature, assuming a single power-law evolution, drops below the condensation temperatures at $\sim10$\,days after the merger.
    }
    \label{fig:cond}
\end{figure}

\section{Condensation temperatures of refractory elements}
\label{sec:mot}
The condensation temperature $T_{\rm c}$ can be defined as the temperature at which the vapor pressure of an element equals its partial pressure. The bottom left panel of Figure~\ref{fig:element} shows $T_{\rm c}$ for each element, calculated from the vapor-pressure data of \citet{Alcock1984} for ejecta with $\Mej=0.05\Msun$ and $\vej=0.1\,c$ at 10\,days after the merger, assuming an ideal gas. The partial pressure of each element is calculated assuming the solar $r$-residual pattern for atomic numbers of $Z>30$ \citep[solid curve in the left top panel of Figure~\ref{fig:element}]{Prantzos2020}. Refractory elements exhibit $T_{\rm c}>1500$\,K, and some of them may be abundantly synthesized by $r$-process.

Figure~\ref{fig:cond} shows the so-called 50\% condensation temperature $T_{50}$ for W ($Z=74$), calculated as follows.
Consider a gaseous element E that condenses into solid state and forms alloy \citep{Palme1976}.
At equilibrium, the standard Gibbs free energy change $\Delta G^o$ is related to the equilibrium constant $K_{\rm eq}$ and the partial pressure $p$ by
\begin{equation}
	\Delta G^o = -k_{\rm B}T \ln K_{\rm eq} = -k_{\rm B}T\ln (a_{\rm E}/p({\rm E})),
	\label{eq:k_eq}
\end{equation}
where $a_{\rm E}$ is the activity of element E in the alloy.
The free energy change is given by $\Delta G^o = \Delta_f G^o(s) - \Delta_f G^o(g)$, where $\Delta_f G^o$ is the standard Gibbs free energy of formation in the solid ($s$) and gas ($g$) phases.
The values of $\Delta_fG^o$ for gaseous and solid W are taken from \citet{JANAF}.
The partial pressures of the condensed element E is set to equal to the partial pressure of the gaseous one:
\begin{equation}
	p({\rm E})=Y({\rm E})p_{\rm tot}(1-\alpha_{\rm E}),
	\label{eq:eq}
\end{equation}
where $p_{\rm tot}$ is the total gas pressure, $Y({\rm E)}$ is the number fraction, and $\alpha_{\rm E}$ is the condensed fraction of element E.
For pure phases, $a_{\rm E}=1$.
For $Y({\rm E)}$, we assume the solar abundance \citep{Lodders2009} or the solar $r$-residuals for $Z>30$ \citep{Prantzos2020}.
The value of $T_{50}$ for pure W is obtained by solving the above equations with $\alpha_{\rm W}=0.5$, assuming an ideal gas with $p_{\rm tot}=n_{\rm tot}k_{\rm B}T$ (dotted and dashed curves in Figure~\ref{fig:cond}).

For alloys, $a_{\rm E}$ becomes
\begin{equation}
	a_{\rm E} = \gamma_{\rm E}\frac{\alpha_{\rm E}Y({\rm E})}{\sum \alpha_i Y(i)},
	\label{eq:a}
\end{equation}
where $\gamma_{\rm E}$ is activity coefficient, and a sum is taken over all elements condensed in the alloy.
Substituting this expression into equation~(\ref{eq:k_eq}) and combining with equation~(\ref{eq:eq}) yields
\begin{equation}
	\frac{1}{\alpha_{\rm E}} = 1 + \frac{\gamma_{\rm E}}{p_{\rm tot}} \frac{1}{\sum \alpha_iY(i)} e^{\Delta G^o/k_{\rm B}T}.
    \label{eq:alloy}
\end{equation}
Here we assume that the third $r$-process peak elements form alloys.
Ideally, one needs to solve a set of coupled non-linear equations for all $\alpha_i$ over all relevant elements, but for simplicity, we assume $\alpha_i = 0.5$ for $i = {\rm W, Re, Os, Ir,}$ and ${\rm Pt}$.
$\gamma_{\rm W}$ is assumed to be unity \citep{Wood2019}.
The resulting $T_{50}$ for alloyed W is shown as the solid line in Figure \ref{fig:cond}.

Figure~\ref{fig:cond} also shows the evolutionary track of kilonova ejecta (see Section \ref{sec:kn_model}).
It is seen that the values of $T_{\rm c}$ and $T_{50}$ for W are consistent with each other at 10\,days.

\section{Formation of dust grains}
\label{sec:form}
As the ejecta expand and cool, the gas temperature decreases below $\sim 1800\,{\rm K}$ at about 10\,days after the merger, providing favorable conditions for dust formation (Figure~\ref{fig:cond}).
To explore whether refractory-element dust can form under the conditions in kilonova ejecta, we solve the time-dependent kinetic rate equations governing the formation of metal clusters in an expanding medium, explicitly including both association and fragmentation (thermal dissociation) processes \citep[see, e.g., ][in the context of supernovae]{Cherchneff2008, Cherchneff2009, Cherchneff2010, Goumans2012MNRAS, Sarangi2013, Lazzati2016ApJ, Sluder2018}. 
Our approach differs from  previous work that concludes that $r$-process grain cannot form because the nucleation timescale exceeds the expansion timescale of the ejecta \citep{Takami2014}.
However, that conclusion relied on the classical nucleation picture, which is not appropriate for the formation of small clusters \citep{Draine1979Ap&SS,Jun2022ARPC}.
A comparison of our results with those obtained using the approach in \citet{Takami2014} is shown in Appendix \ref{sec:comparison}.

\subsection{Kinetic Equation}
\label{sec:kinetic}
We consider that refractory elements condense together to form alloy grains. The evolution of the number density of $n$-mers, $n_n$, is explicitly followed by solving a set of $N$ nonlinear differential equations, where $N$ is the maximum cluster size.
We include 
(i) association: W$_m$ + W$_{n-m}$ $\rightarrow$ W$_n$,
(ii) fragmentation (thermal dissociation): W$_{n}$ $\rightarrow$ W$_{m}$ + W$_{n-m}$, and
(iii) destruction by fast electrons produced by $\beta$-decay: W$_{n}$ $+$ $e^-\rightarrow$ W$_{m}$ + W$_{n-m}$.
The general equation for the evolution of $n$-mers in expanding matter is given by 
\begin{equation}
\begin{aligned}
	\frac{dn_n}{dt} = &-\sum_{m=1}^{n/2} k^{\rm frag}_{m,n} n_n
				   +\sum_{m=1}^{n/2} k^{\rm assoc}_{m,n} n_m n_{n-m} \\
				  & -\sum_{m=1}^{n/2} k_{\beta,mn}n_n 
				 +\sum_{j=n+1}^{N} s_{jn} k^{\rm frag}_{n,j} n_j \\
				  & -\sum_{j=n+1}^{N} s_{jn} k^{\rm assoc}_{n,j} n_{j-n} n_n
				 +\sum_{j=n+1}^{N} s_{jn}k_{\beta,nj}n_j \\
				  &- \frac{3n_n}{t},
	\label{eq:rate_eq}
\end{aligned}
\end{equation}
where $k^{\rm assoc}_{m,n}$ and $k^{\rm frag}_{m,n}$ are the reaction rates of (i) association and (ii) fragmentation, respectively, and $s_{ij}$ is the stoichiometric coefficient equal to 2 if $j-i=i$ otherwise 1. $k_{\beta, mn}$ is the rate coefficient for the process (iii) that non-thermal $\beta$-electrons hit and dissociate the formed clusters in the ejecta, and the last term corresponds to the decrease in number density due to homologous expansion.

The reaction rate coefficients for all condensed elements are approximated by those of W clusters, extrapolated from the results of a molecular dynamics simulation \citep{Wcluster} except for the dimer formation rate, $\kdi=k^{\rm assoc}_{1,2}$.
Although \citet{Wcluster} found $\kdi\sim 10^{-10}\,{\rm cm^3\,s^{-1}}$, this value is likely overestimated \citep{Li2008}.
Radiative association is typically inefficient, and thus $\kdi$ is expected to be much lower than those of larger clusters. 
For example, radiative association rate coefficients for molecules such as C$_2$, SiO, SiS, and TiO are typically $\sim10^{-17}$--$10^{-16}$\,cm$^3$\,s$^{-1}$ \citep[e.g.][]{Clayton1999, Clayton2001,Andreazza2007, Andreazza2012, Babb2019ApJ,Hou2023A&A}. Of course, the actual dimer formation rates for refractory metals could be even smaller than this range. However, the average of known radiative association rates for light-element species is roughly of this order, and the TiO formation rate represents a plausible upper bound for these processes of $d$-shell elements.
Therefore, although the actual dimer formation rate is uncertain, we adopt $\kdi=10^{-17}\,{\rm cm^3\,s^{-1}}$ as a representative baseline value.

For $k_{\beta, mn}$, we adopt
\begin{equation}
\begin{aligned}
    k_{\beta,mn} &= \int y(t,E) \sigma_{{\rm EI},mn}(E) dE \\
    &=\frac{f_e\dot{q}_\beta(t)\rho}{E_\beta}\int\frac{1}{n_{\rm gas}\sigma_{\rm st}(E)}\sigma_{{\rm imp},mn}(E)dE,
     \label{eq:kb}
\end{aligned}
\end{equation}
where $\sigma_{\rm EI}(E)$ is the electron-impact dissociative (partial) ionization cross section for clusters, and the integration is performed over 5\,eV--300\,keV.
The quantity
\begin{equation}
	y(t, E) = \frac{S_\beta(t)}{n_{\rm gas}\sigma_{\rm st}(E)} 
    = \frac{f_e\dot{q}_\beta(t)\rho}{E_\beta}\frac{1}{n_{\rm gas}\sigma_{\rm st}(E)}
    \label{eq:y_def}
\end{equation}
is the number flux density per unit energy of $\beta$-electrons in the ejecta. Here $S_\beta(t)$ is the production rate of $\beta$-electrons per volume, $\dot{q}_\beta(t)=2\times10^{10}(t/{\rm day})^{-1.3}\,{\rm erg\,s^{-1}\,g^{-1}}$ is the specific heating rate by $\beta$-decays, with $f_e=0.25$ the fraction carried by electrons \citep[e.g.,][]{Hotokezaka2020}, and $\rho$ is the ejecta density.
We adopt $E_\beta=300$\,keV for the injection energy of primary $\beta$-electrons.
The gas number density is $n_{\rm gas}=\rho/Am_u$, where $A$ is the mass number and $m_u$ is the atomic mass unit.
The stopping cross section is defined as $\sigma_{\rm st}(E)=K_{\rm st}(E)\cdot Am_u$, where $K_{\rm st}$ is the stopping power per unit mass for electrons by ionization/excitation of atomic gas.
We assume $K_{\rm st}\sim1(v/c)^{-1}$ in units of ${\rm MeV\,cm^2\,g^{-1}}$ \citep{Waxman2018,Hotokezaka2020}.
For large gas density or large stopping cross section, $\beta$-electrons lose energy efficiently, reducing the flux of high-energy electrons $y$ that may interact with clusters.
Substituting these expressions, equation~(\ref{eq:y_def}) becomes
\begin{equation}
\begin{aligned}
	y(t, E) &= \frac{f_e \dot{q}_\beta(t)}{E_\beta K_{\rm st}(E)} \\
		      &\approx 0.25\times10^{16} \left(\frac{t}{\rm 1\,day}\right)^{-1.3} \frac{v(E)}{c} \frac{1}{E_\beta}\,{\rm cm^{-2}\,s^{-1}\,eV^{-1}}.
	\label{eq:y}    
\end{aligned}
\end{equation}

In this work, we consider destruction by $\beta$-electrons only for dimers, \ie $k_\beta=k_{\beta,21}$. This is motivated by the fact that dimer formation can be slow and $\beta$-electrons may slow down the grain formation process.
To our knowledge, data for $\sigma_{\rm EI}(E)$ for refractory-element dimers are unavailable. Instead, we use the theoretical data for WO molecules from \citet[adopted BEB results; see their Table 3]{Wbeta}, assuming similar behavior for metal refractory-element dimers.
While cross sections for larger clusters are unknown, those for complex molecules such as WF$_6$, hydrocarbon, and alcohol are comparable to or smaller than those of WO \citep[e.g.,][]{hydrocarbon,alcohol,WF6}.
This suggests that $\beta$-electron destruction may be less important for larger clusters (but see Section~\ref{sec:radioactive}).

Finally, the maximum cluster size is set to $N=20$ throughout this work, limited by the available rate coefficients. In practice, grains are expected to grow to much larger sizes (see Section~\ref{sec:res}).

We note that there are additional processes that could cause grain destruction.
First, we do not consider destruction of grains by photons.
High-energy photons may interact with grains, causing them to be charged or even destructed \citep{Weingartner2001, Weingartner2006}.
However, although $\beta$-decay produces both high-energy electrons and $\gamma$-rays, most $\gamma$-rays escape without significant interaction after a few days after the merger due to the low opacity \citep{Barnes2016,Hotokezaka2016,Guttman2024MNRAS}.
Moreover, thermal UV photons are not important as the ejecta temperature is $\lesssim 5000$\,K after a few days after the merger (see Section~\ref{sec:kn_model}).

We also neglect destruction of grains by $\alpha$-particles and fission fragments, effectively assuming that heating is dominated by $\beta$-decay.
Because heating rates from $\alpha$-decay and spontaneous, neutron-induced, or $\beta$-delayed fission are largely uncertain depending on the elemental abundances as well as nuclear mass models \citep[e.g.,][]{Wu2019, Barnes2021, Zhu2021}, this assumption provides a reasonable baseline. However, if energetic $\alpha$-particles or fission fragments significantly contribute to the energy deposition, they could impact dust grain formation and survival. We discuss such potential scenarios in Section~\ref{sec:radioactive}.

Radioactive isotopes with half-lives of $\sim10$--50\,days could trigger in-grain destruction on the timescales of interest. However, among the key elements under consideration such as Zr, Mo, W, Re, Os, and Ir, only $^{191}$Os has a half-life in this range, $\sim15$\,days, and most other isotopes have much shorter or longer half-lives. While a few isotopes such as $^{95}$Zr (half-life of $\sim64$\,days) and $^{185}$W ($\sim75$\,days) may be relevant for dust survival, these specific isotopes constitute only a small fraction of the abundance among refractory elements. Thus, we neglect in-grain destruction in the present calculations.

As can be found in the supernova literature, the chemical and molecular network is considerably more complex than the one considered here \citep[e.g.,][]{Cherchneff2008, Cherchneff2009, Cherchneff2010, Sarangi2013, Lazzati2016ApJ, Sluder2018}. It can typically involve recombination, three-body interactions, charge exchange, and chemical weathering acting as either formation and destruction processes. We omit most of these processes here, primarily due to the lack of established reaction rates for metal molecules and clusters. Furthermore, a rigorous treatment would require coupling the kinetic equations directly to the ionization and temperature structures of the ejecta to track gas and dust evolution simultaneously (see also Section \ref{sec:kn_model}). As a first approximation, we adopt a simplified approach that sets aside these complexities. As demonstrated in Appendix \ref{sec:comparison}, while our framework can reproduce the results of \citet{Takami2014} when configured to mimic their setup, it will reveal fundamentally different physical behavior under its native application.

\subsection{Timescales}
\label{sec:timescale}
The timescales of the processes considered in equation~(\ref{eq:rate_eq}) are useful for understanding the results.
The first is the expansion timescale:
\begin{align}
    t_{\rm exp} = |(1/\rho)(d\rho/dt)|^{-1}.
\end{align}
As described below, we assume uniform ejecta with homologous expansion, which gives $t_{\rm exp}=t/3$, where $t$ is time since the merger.

The second is the association timescale. For dimers,
\begin{equation}
\begin{aligned}
    t_{\rm di} & =[\kdi n_{\rm ref}]^{-1}\\
               & = 10^{11} \bigg(\frac{\kdi}{10^{-17}\,{\rm cm^{3}\,s^{-1}}}\bigg)^{-1}
               \bigg(\frac{n_{\rm ref}}{10^{6}\,{\rm cm^{-3}}}\bigg)^{-1} \,{\rm s},
\end{aligned}
\end{equation}
where $n_{\rm ref}$ is the number density of condensable refractory elements.
On the other hand, for the reaction W$_1$ + W$_{n-1}$ $\rightarrow$ W$_n$, the rate coefficients $k^{\rm assoc}_{1,n}$ are of order $10^{-10}$\,cm$^3$\,s$^{-1}$ at $\sim 2000\,{\rm K}$ \citep{Wcluster}.
Thus, the association timescale for large clusters is approximately $t^{\rm assoc}_{1,n} = [k^{\rm assoc}_{1,n}n_{\rm ref}]^{-1}\sim 10^{10}n_{\rm ref}^{-1}$\,s.
A necessary condition for dust formation is that the association timescale is shorter than the expansion timescale.

The association timescale can be compared with the collisional timescale:
\begin{equation}
	t_{\rm coll} = \left[4\pi a_1^2 n_{\rm ref} \left(\frac{k_BT}{2\pi m_{\rm ref}}\right)^{1/2}\right]^{-1},
	\label{eq:tcoll}
\end{equation}
where $a_1$ denotes the monomer (atomic) radius of the species that compose grains, $m_{\rm ref}$ is the atomic mass of the refractory elements, and $T$ is the ejecta temperature.
For an atom with $A=190$, the collisional timescale is
\begin{equation}
	t_{\rm coll} \approx 2.9\times 10^{4} \bigg(\frac{n_{\rm ref}}{10^{6}\,{\rm cm^{-3}}}\bigg)^{-1} 
	\bigg(\frac{a_1}{1.56\,\textrm{\AA}}\bigg)^{-2} \bigg(\frac{T}{1800\,{\rm K}}\bigg)^{-1/2}~{\rm s}.
\end{equation}
This is consistent with the association timescale for large clusters.

Next, the fragmentation (thermal dissociation) timescale is $t^{\rm frag}_{m,n}= [k^{\rm frag}_{m,n}]^{-1}$.
The value of $k^{\rm frag}_{m,n}$ depends on the cluster size;
for example, $k^{\rm frag}_{1,n}$ ranges from $\sim 10^{-12}$ to $10^{-2}$\,s$^{-1}$ at the temperature of interest \citep{Wcluster}.

Finally, the destruction timescale of dimers by $\beta$-electrons is
\begin{align}
    t_\beta &= k_\beta^{-1} \approx 5.0\times10^4 \left(\frac{t}{\rm 10\,day}\right)^{1.3}\,{\rm s}.
     \label{eq:tb}
\end{align}

\subsection{Kilonova Model}
\label{sec:kn_model}
We consider homologously expanding ejecta with uniform density $\rho(t)$ and temperature $T(t)$. The density is given by $\rho=3\Mej/4\pi(\vej t)^3$, where $\Mej$ and $\vej$ are the ejecta mass and velocity. The average number density $n_{\rm tot}$ is then obtained by adopting the average atomic mass of $\bar{A}=130$. We assume the mass fraction of condensable refractory elements of $X_{\rm ref}=0.034$, corresponding to the sum of the mass fractions of W, Re, Os, Ir, and Pt for the solar $r$-residuals for $Z>30$ \citep{Prantzos2020}.

The temperature is approximated by a single power law, 
\begin{equation}
	T(t) = T_0 \left(\frac{t}{t_0}\right)^{-\gamma},
	\label{eq:T}
\end{equation}
where we adopt $T_0=2500$\,K, $t_0=6$\,days, and $\gamma=0.565$, empirically obtained by fitting to the blackbody temperatures of the spectra of AT2017gfo at $t=0.48$--10.4\,days \citep[and references therein]{Hotokezaka2023, Sneppen2024}.

We note that the actual temperature evolution of the ejecta is not trivial.
The gas temperature decreases until the ejecta enters the nebular phase, where the temperature is determined by the thermal balance between radioactive heating and radiative cooling. 
According to non-local thermodynamic equilibrium (non-LTE) calculations, the gas temperature begins to slowly increase with time in the nebular phase if atomic processes are only cooling sources \citep{Hotokezaka2021, Pognan2022,Brethauer2025}. 
For example, \citet{Hotokezaka2021} showed that the pure-Nd gas temperature cannot easily decrease to $\sim2000$\,K. However, atomic cooling rates strongly depend on the underlying atomic structure. In reality, the ejecta contains a complex mixture of heavy elements whose transitions could collectively enhance radiative cooling, allowing the gas to cool much more efficiently. Computing such multi-element cooling functions requires comprehensive atomic data, including recombination rates and collisional strengths, which are currently unavailable.

Observations of AT2017gfo provide a possible indication of substantial cooling. Its SED shifted toward longer wavelengths with time, with the relative contribution of optical emission decreasing significantly compared to the infrared at later epochs \citep{Pian2017, Smartt2017}. The apparent continuum temperature dropped below 2000\,K by $\sim 10$\,days after the merger (\citealp{Hotokezaka2023}; but see also, \eg\citealp{Waxman2018}). We emphasize that the electron temperature does not necessarily track the radiation temperature, as the two can decouple during the nebular phase. However, high electron temperatures $\gg 2000\,{\rm K}$ would excite optical emission lines, which are very weak in the 10.4-day spectrum. This may indicate that the electron temperature is relatively low. Nevertheless, line opacity could also reprocess optical flux into the infrared, resulting in an apparently cool SED even if the electron temperature remains relatively high.

The possible thermal state of kilonova ejecta may differ from Type Ia supernovae. In Type Ia supernovae, nebular spectra exhibit strong optical emission lines of Fe-group elements, indicating electron temperatures of $\gtrsim 5000\,{\rm K}$, with no evidence of dust formation \citep[e.g.,][]{Liu1997, Mazzali2011, Kwok2023}.
In contrast, kilonova ejecta contain heavy elements, particularly lanthanides, whose rich atomic structures and numerous transitions may provide more efficient radiative cooling. This could allow kilonova ejecta to reach lower temperatures than those typically inferred for Type Ia supernovae.

In summary, we adopt a power-law temperature evolution as a purely empirical prescription, rather than as a prediction from self-consistent thermal-balance calculations. Although a low electron temperature has not yet been firmly established observationally or theoretically, the available data do not rule out such a cool thermal state.
Our primary motivation is therefore to investigate the physical consequences under the assumption that the ejecta reach low temperatures. We will address the electron temperature of kilonovae in future work.

The ionization structure of the ejecta also plays a crucial role in dust formation. While our kinetic calculations assume a fully neutral gas (100\% neutral fraction), early cluster growth could in principle involve both neutral-neutral and neutral-ion interactions.
Although dense inner regions maintain substantial neutral fractions via rapid recombination \citep[e.g.,][]{Pognan2022, Pognan2026, Brethauer2025}, charging introduces destructive channels: Coulomb repulsion prevents ion-ion dimer association, while small molecular ions can undergo rapid dissociative recombination.
For example, models by \citet{Brethauer2025} showed a neutral gas fraction of $\sim 10\%$ at 5--10\,days, effectively reducing the neutral-neutral dimer formation rate by two orders of magnitude. Because key reaction rate coefficients, including dissociative recombination, neutral-ion association, and charge transfer for $r$-process elements, are currently unavailable, it remains unclear whether charging net-suppresses or alters early cluster formation. 
We thus regard the uncertainty of the ionization state into the effective $\kdi$; how the resulting dust mass vary under different choice of $\kdi$ is demonstrated in Appendix~\ref{sec:k_di}.
Overall, while ionization clearly suppresses early small cluster formation, its net impact on subsequent growth into larger dust grains remains uncertain. 
Fully resolving the competition between these charged channels requires non-LTE plasma calculations coupled with charged reaction networks, which we leave to future work.

\begin{figure*}
    \centering
    \includegraphics[width=\linewidth]{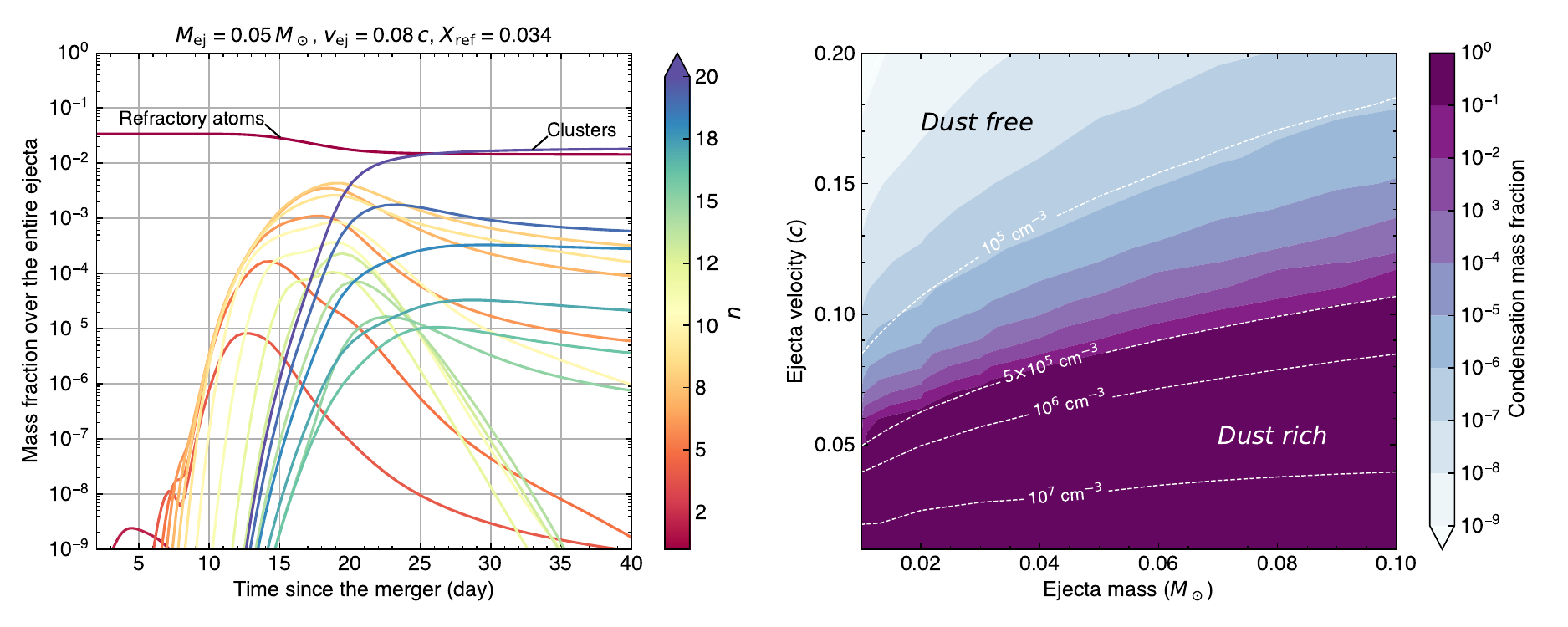}
    \caption{
    Left: evolution of the mass fraction of clusters in the ejecta for $\Mej=0.05\,\Msun$ and $\vej=0.08\,c$, with the condensable refractory-element mass fraction of $X_{\rm ref}\approx 3$\%. Different colors indicate clusters of different sizes as shown in the color bar.
    Right: condensation mass fraction, defined as the ratio of the dust mass $\Md$ to the total mass of refractory elements $M_{\rm ref}$, as a function of ejecta mass and velocity. The condensable refractory-element mass fraction is $\approx3$\%, such that $M_{\rm ref}\approx0.03\times\Mej$. The temperature evolution is assumed to be identical for all ejecta models. The dashed contours show the number density of the refractory elements in uniform-density ejecta at $\approx9.6\,$days when the ejecta temperature drops to 2000\,K.
    }
    \label{fig:abun}
\end{figure*}

\subsection{Results of Dust Formation}
\label{sec:res}
The resulting cluster formation history for $\Mej=0.08\,\Msun$ and $\vej=0.08\,c$ is shown in Figure~\ref{fig:abun}. 
At early times, dimers form slowly, while high temperatures suppress the growth of larger clusters. As the temperature decreases and the abundance of larger clusters become comparable to that of dimers, larger clusters begin to grow rapidly. 
The ejecta continue to expand, and the association rates eventually fall below the expansion rate, causing the cluster abundances to freeze out. 
Dimer formation acts as the bottleneck controlling dust formation efficiency, with the dimer abundance determined by both temperature and density (see below).

Although our simulation explicitly follows cluster growth only up to $n=20$, further growth to larger sizes is expected.
The characteristic grain size, $a_c$, can be estimated by comparing $t_{\rm coll}$ with $t_{\rm exp}$ \citep[see Section \ref{sec:timescale};][]{Todini2001, Nozawa2003}, yielding 
\begin{equation}
\begin{aligned}
    a_c & \approx a_1\left(\frac{t_{\rm exp}}{t_{\rm coll}}\right),\\
    & \approx 5\,{\rm nm}\,\bigg(\frac{n_{\rm ref}}{10^{6}\,{\rm cm^{-3}}}\bigg)\left(\frac{T}{1800\,{\rm K}}\right)^{1/2}\left(\frac{t_{\rm exp}}{10\,{\rm day}}\right),
    \label{eq:a_c}
\end{aligned}
\end{equation}
where $n_{\rm ref}$ is the atomic number density of refractory elements during the dust formation phase and $a_1\approx 1.56\,\textrm{\AA}$ is the monomer radius.
The number of atoms in a grain is then
\begin{equation}
\begin{aligned} 
	N &\approx \frac{4\pi}{3}a_1^3\left(\frac{t_{\rm exp}}{t_{\rm coll}}\right)^3\rho_{\rm d} (Am_u)^{-1} \\
	    &\approx 3.2\times 10^3\\
	    & \quad \bigg(\frac{\Mej}{0.05\,\Msun}\bigg)^3 \bigg(\frac{\vej}{0.05\,c}\bigg)^{-9} \bigg(\frac{t}{10.7\,{\rm day}}\bigg)^{-6} \\
	                 &  \quad\bigg(\frac{X_{\rm ref}}{0.034}\bigg)^3 \bigg(\frac{T}{1800\,{\rm K}}\bigg)^{-3/2}.
	\label{eq:N}
\end{aligned}
\end{equation}

Based on these considerations, clusters reaching $n = 20$ are expected to grow to larger sizes. 
We therefore estimate a final dust mass after freeze-out of $10^{-3}\,\Msun$ based on the mass fraction of clusters with $n=20$.
We note that the kinetic cluster formation from the gas phase governs the nucleation phase, while dust grains eventually form via coalescence and coagulation of clusters with each other in the condensation phase; the latter is not considered in the present work \citep{Sarangi2013}.
Thus, the derived dust mass should be regarded as an upper limit, but investigating the simultaneous condensation of a mixture of metal elements remains a non-trivial task and will be a subject of future work.

The efficiency of dust formation depends sensitively on the physical conditions of the ejecta. 
The dependence of the final mass fraction of $r$-process dust on ejecta velocity and mass is shown in the right panel of Figure~\ref{fig:abun}. Dust formation is efficient in slower components, with velocities of $\lesssim 0.1\,c$, whereas faster components with $\gtrsim 0.1\,c$ are expected to remain largely dust-free. Notably, the expansion velocity inferred from the photospheric radius of the infrared continuum emission in AT2017gfo and AT2023vfi is $\lesssim 0.1\,c$, providing observational support for this velocity threshold for dust formation.

We note that the final dust mass presented here adopts the baseline value of $\kdi=10^{-17}$\,cm$^3$\,s$^{-1}$. Efficient dust formation requires exceeding a critical density at the condensation epoch $t_0$, when $T$ drops below $T_{\rm c}$. As shown by our analytic formalisms in Section~\ref{sec:analytic}, this threshold corresponds to $t_0>t_a\ln(t_d/t_a)$, where $t_a\propto1/\rho$ and $t_d\propto1/\kdi$ are the key formation timescales. Reducing $\kdi$ increases $t_d$, pushing low-density regions below this critical threshold and strongly suppressing dust formation, which leads to an almost linear scaling of dust mass with $\kdi$. In contrast, for high-density regimes that remain comfortably above this threshold, exponential cluster growth is successfully sustained, and the final dust mass depends only weakly on $\kdi$ (see Figure~\ref{fig:k_di} in Appendix~\ref{sec:k_di}). Thus, while $\kdi$ remains uncertain, the picture of rapid dust formation robustly holds across high-density ejecta.

Furthermore, Figure~\ref{fig:abun} assumes uniform ejecta density and identical temperature evolution for a given $\Mej$ and $\vej$. The exact boundary for efficient dust formation depends on $t_0$ for a given density. In a realistic ejecta profile with density and temperature gradients, the outer high-velocity layers are less dense but also cooler than the inner ejecta. The earlier condensation epoch $t_0$ can allow the outer layers to satisfy the condition for efficient dust condensation, as seen in Section~\ref{sec:rad}.

\begin{figure*}
    \centering
    \begin{tabular}{cc}
        \includegraphics[width=0.45\linewidth]{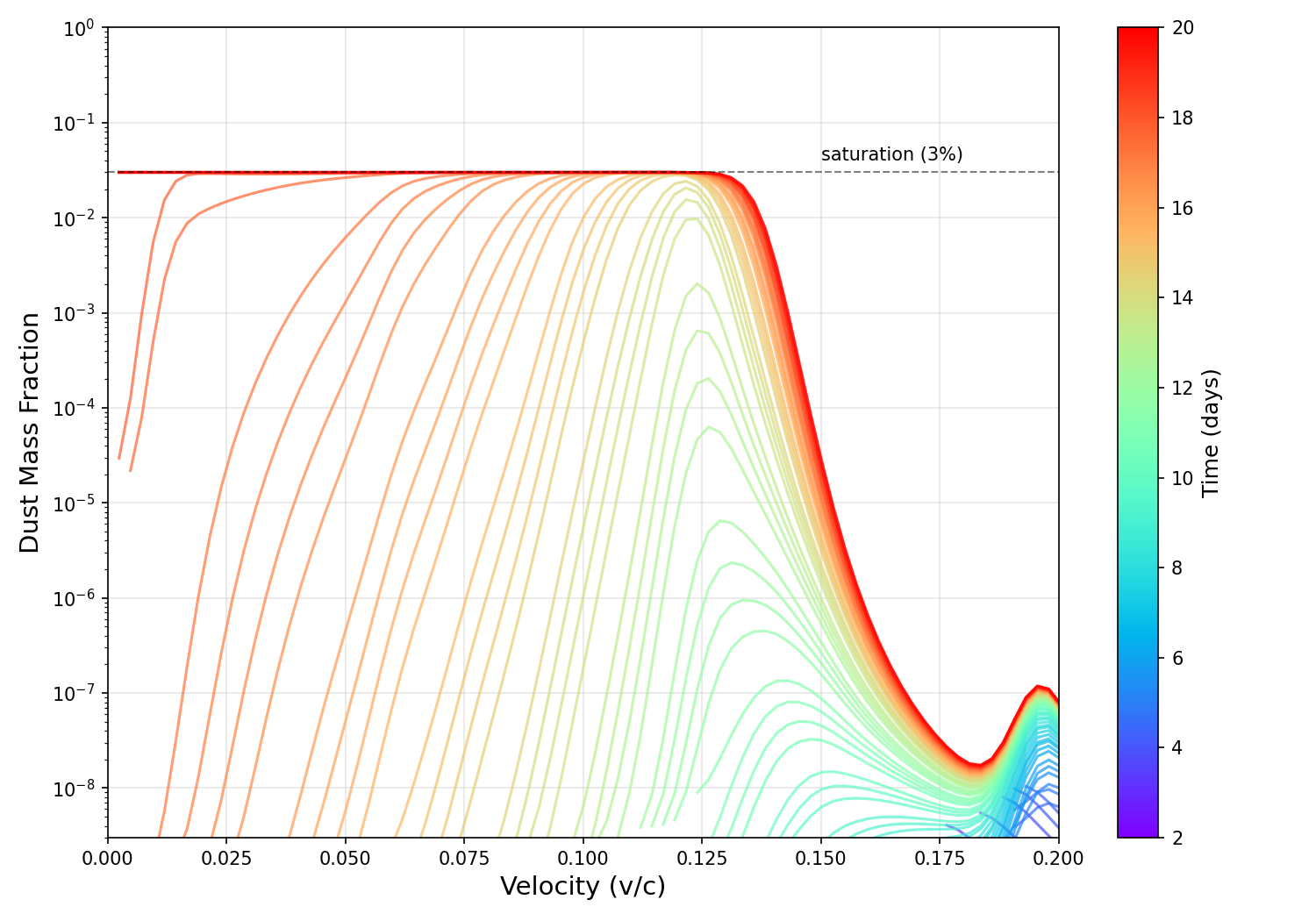}
        \includegraphics[width=0.55\linewidth]{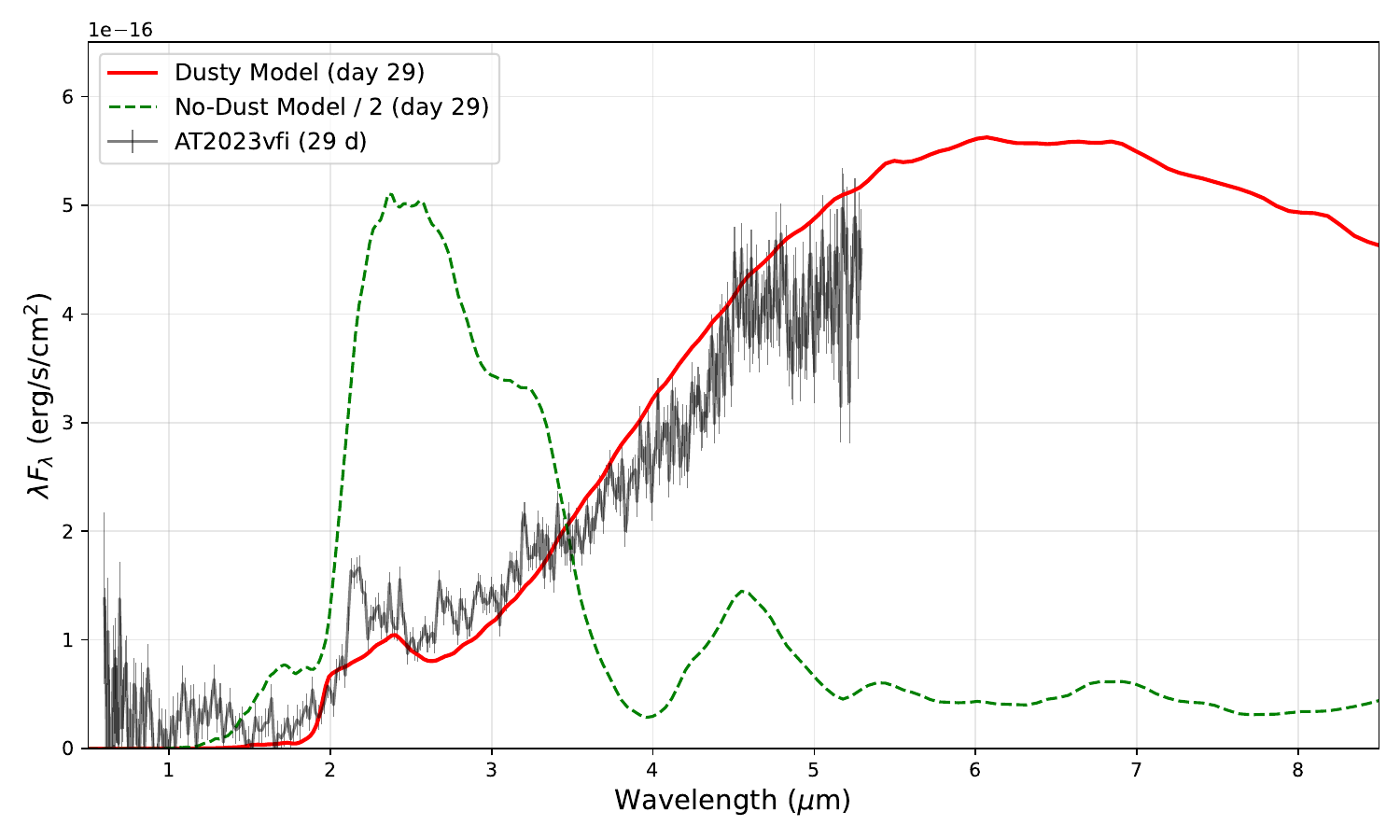}
    \end{tabular}
    \caption{
    Left: dust mass fraction as a function of ejecta velocity for $\Mej=0.1\,\Msun$ in the radiative transfer calculation. Different colors indicate the temporal evolution. The dust mass fraction saturates at 3\%, corresponding to the assumed mass fraction of condensable refractory elements.
    Right: comparison of the model spectrum of a dusty kilonova (red solid curve) with the observed spectrum of AT2023vfi at 29\,days \citep{Gillanders2025}. The dashed curve shows the same model but without dust (see also Figure \ref{fig:spectrum_dust_nodust}).
    }
    \label{fig:spectrum}
\end{figure*}

\section{Radiative transfer modeling of dusty kilonovae}
\label{sec:rad}
\subsection{Methods}
To simulate the emission from dusty kilonova ejecta we carry out time-dependent multi-wavelength radiation transport calculations using the Sedona code \citep{Kasen2006}.
The assumed kilonova ejecta structure is analogous to that described in \citet{Brethauer2025}, a broken power law with a shallow profile ($\rho \propto r^{-1}$) in the inner layers and a steep cutoff ($\rho \propto r^{-8}$) in the outer layers. The transition between these regions occurs at a velocity coordinate $v_{\rm t}=0.12\,c$.
The ejecta is taken to be spherically symmetric and homologously expanding, with a uniform composition of heavy $r$-process elements in a solar abundance pattern. 

The kilonova emission is powered by radioactive heating from the decay of $r$-process isotopes. The rate of radioactive energy emission and the efficiency by which the decay products are thermalized in the ejecta follow those described in \citet{Brethauer2025}. The temperature of the ejecta in each zone and at each time step is calculated self-consistently by balancing the cooling by thermal emission with the heating by radioactivity and the absorption of radiation. 

The opacity and emissivity of the ejecta are calculated by adopting LTE in each zone to determine the ionization and excitation states. The radiative processes included are bound-free, free-free, and bound-bound transitions, as well as dust absorption/emission. In kilonova ejecta, the forest of bound-bound line transitions -- in particular those of the lanthanides and actinides -- are Doppler broadened by the high ejecta velocities into a pseudo-continuum  that typically dominates the opacity. We use an extensive database of atomic lines and levels derived from the atomic structure models of \citet{Tanaka2020}.

To incorporate dust formation in the ejecta into radiative transfer calculations, we calculate a dust mass fraction $X_{\rm d}$ in each zone and determine its associate opacity and emission. Modeling a dust grain as a sphere of radius $a$ and of uniform density $\rho_{\rm d}$, the mass of a grain is $m_{\rm d} = (4\pi/3) \rho_{\rm d} a^3$ and so the number density of grains is 
\begin{equation}
    n_{\rm d} = \frac{X_{\rm d}~\rho}{m_{\rm d}} = \frac{3}{4\pi} \frac{X_{\rm d}~\rho}{\rho_{\rm d} a^3}~.
\end{equation}
The absorptive cross-section of a dust grain of size $a$ is  $\sigma_{\rm d,abs}(\lambda) = Q_{\rm abs}(\lambda) \pi a^2$ where $Q$ is the efficiency coefficient. We predict nanometer sized grains, and so adopt the Rayleigh limit ($a \ll \lambda$) value, $Q_{\rm abs} \approx a/\lambda$. The scattering efficiency  in the Rayleigh limit, $Q_{\rm sc} \approx (a/\lambda)^4$, is much smaller, so the dust grains in the infrared are nearly completely absorbing and $\sigma_{\rm d}(\lambda) \approx \pi a^3/\lambda$. The opacity (units ${\rm cm^2~g^{-1}}$) of dust is then 
\begin{equation}
    \kappa_{\rm d} = \frac{n_{\rm d}\sigma_{\rm d,abs}}{\rho} = \frac{n_{\rm d}}{\rho} \frac{\pi a^3}{\lambda} 
               = \frac{3}{4} \frac{X_{\rm d}}{\rho_{\rm d}} \frac{1}{\lambda}~. 
\end{equation}
In the Rayleigh limit, the grain size cancels out so the distribution of grain sizes does not influence the opacity (see also Section~\ref{sec:mass}). 

The emissivity of the dust is assumed to be thermal, and so by Kirchhoff's law is $\epsilon_{\rm d} = \rho \kappa_{\rm d} B_\nu(T)$, where $B_\nu(T)$ is the Planck function. The dust temperature is assumed to be in thermal equilibrium with the gas, and radiative heating and cooling by both dust and gas are included in our solution for temperature. In regions of significant dust formation, the dust dominates the heating and cooling and so primarily determines our calculated temperature. If the dust-gas coupling is weak, our calculated temperature should  properly reflect the thermal state of the dust, but the temperature of the gas may differ significantly.

\subsubsection{Analytic Model for Dust Formation}\label{sec:analytic}
We calculate the dust mass fraction, $X_{\rm d}$, as a function of time in each zone using a simplified analytic prescription for grain growth that captures the key aspects of the detailed kinetics studied in Section~\ref{sec:form}. Dust grains are only assumed to survive when the temperature is less than a condensation temperature chosen to be $T_{\rm c} = 1600$\,K.  If $f$ is the fraction of atoms of a particular species that are locked into clusters or dimers, we model the rate at which this fraction grows as
\begin{equation}
   n  \frac{df}{dt} = k^{\rm assoc}_{1,n}  ~ n_g ~n_m
  + 2 k_{\rm di}^{\rm assoc} n_m^2
  \label{eq:df_dt_transport}
\end{equation}
where $n$ is the total number density of the species and $n_m = (1-f) n$ is the density of monomers. For simplicity, we take each cluster to be composed of a characteristic number of atoms, $\bar{N}_g$, such that the number density of clusters is $n_{g} = n f/\bar{N}_g$. The first term on the right-hand side of equation~(\ref{eq:df_dt_transport}) represents the association of a monomer onto a cluster (with rate coefficient $k^{\rm assoc}_{1,n}$) and the second term represents the association of two monomers into a dimer (with rate coefficient $k_{\rm di}^{\rm assoc}$).
For this analytic prescription, the effects of $\beta$-electrons are omitted. While this simplification may slightly overestimate the final dust mass, it does not change the underlying physical behavior of the model and does not impact our global discussions.

For homologously expanding ejecta, the number density of the species evolves as $n(t) = n_{0} (t_{0}/t)^3$ which we have scaled to the recorded density, $n_{0}$, and time, $t_{0}$, at which the temperature in a zone last dropped below $T_c$.
The solution to equation~(\ref{eq:df_dt_transport}) with the initial condition $f(t_0) = 0$ is then (for times $t \geq t_0$)
\begin{equation}
    f(t) = \frac{\xi(t)}{1 + \xi(t)} ~~{\rm where}~
    \xi(t) =  \frac{t_a}{t_d} \left[ e^{\frac{t_0}{t_a}\left( 1 - \frac{t_0^2}{t^2} \right)}   - 1 \right]
    \label{eq:f_dust_growth}
\end{equation} 
where $t_d = 1/(k_{\rm di}^{\rm assoc} n_0)$ and $t_a = 2 \bar{N}_g/(k^{\rm assoc}_{1,n} n_{0})$ are characteristic timescales for atoms to associate with monomers and clusters, respectively, at time $t_0$ (cf.~Section \ref{sec:timescale}). At early times, the fraction evolves as $f(t) \approx 2 (t- t_0) /t_d$ , reflecting how the slow process of dimer formation initially acts as a bottleneck to exponential cluster growth. 
The condition for a species to effectively form dust ($f(\infty) \approx 1$) is $ t_a < t_0/ \ln(t_d/t_a)$, where for typical values $\ln(t_d/t_a) \approx 10$.   

We use the same parameters to describe the grain growth of all condensing species, $k_{\rm di}^{\rm assoc} \approx 10^{-17}\,{\rm cm^3~s^{-1}}$, $k^{\rm assoc}_{1,n} \approx 10^{-10}{\,\rm cm^3~s^{-1}}, \bar{N}_g = 100$. The total dust mass fraction in a zone is then the sum over all refractory elements in the ejecta
\begin{equation}
    X_{\rm d}(t) = \bar{A}^{-1} \sum_i f_{i}(t)~Y_i ~A_i 
\end{equation}
where $A_i$ is the atomic mass of species $i$ and $\bar{A}$ is the average atomic mass. 

While this analytic treatment is approximate, it reproduces the general behavior of the detailed kinetic calculations of Figure~\ref{fig:abun}. Because the cluster growth is exponential once it gets started, saturation is the likely outcome whenever the criterion for dust formation is met in a zone. The primary effects of dust on the radiative transfer are therefore relatively insensitive to the detailed time evolution.

\begin{figure}
    \centering
	\includegraphics[width=\linewidth]{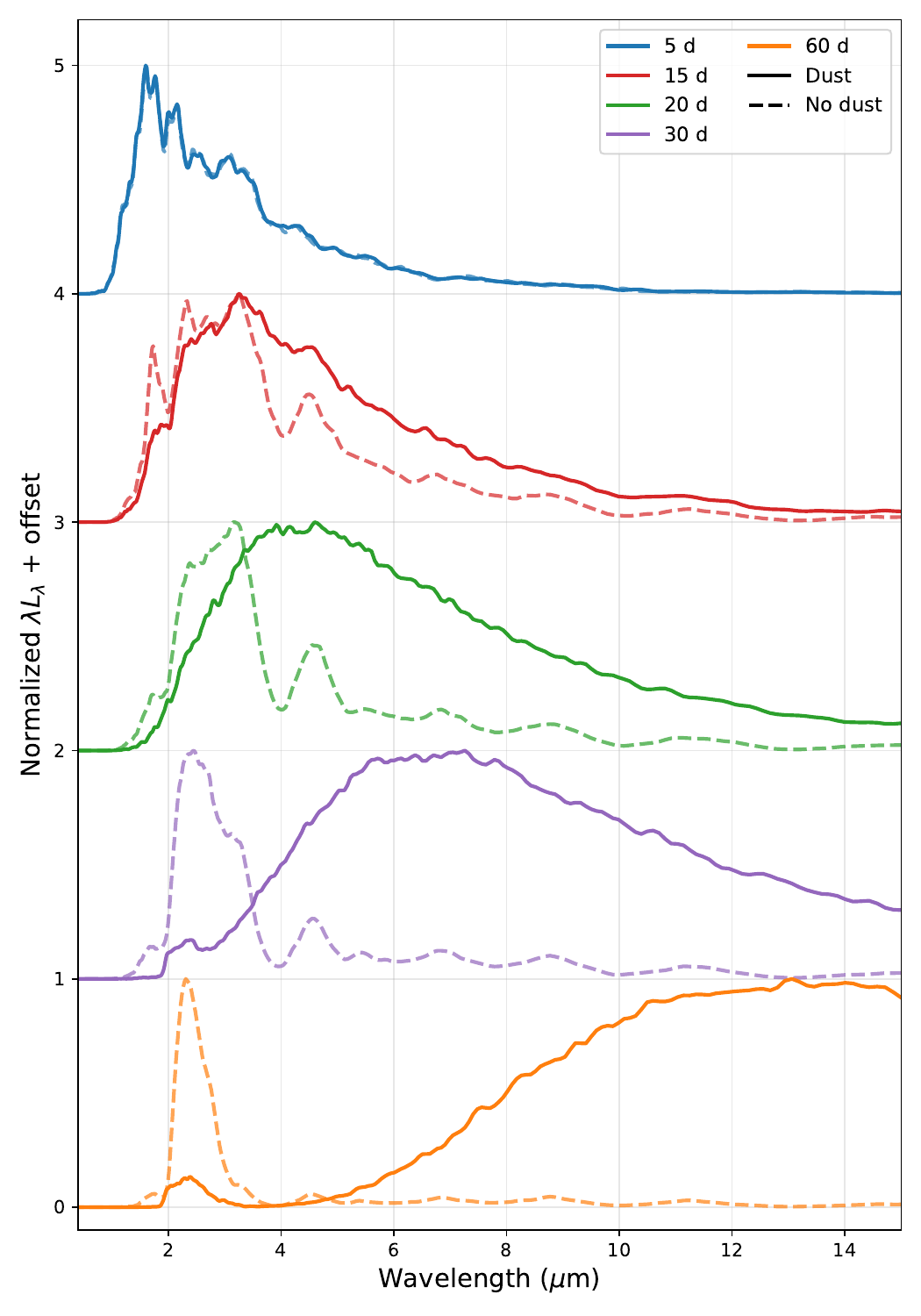}
    \caption{
   Synthetic spectra from the kilonova radiative transfer calculations with and without the inclusion of dust. When no dust is included,  the ejecta at later times ($\gtrsim 20$\,day) is optically thin and the luminosity emerges from complexes of line emission, most prominently those near $2\,\mu$m. When dust is included, the dusty inner ejecta produce optically thick blackbody emission, while the dust-free outer layers produce a smaller component of optically thin emission visible as the peak near $2\,\mu$m.
    \label{fig:spectrum_dust_nodust}
    }
\end{figure}

\begin{figure*}
	\includegraphics[width=\linewidth]{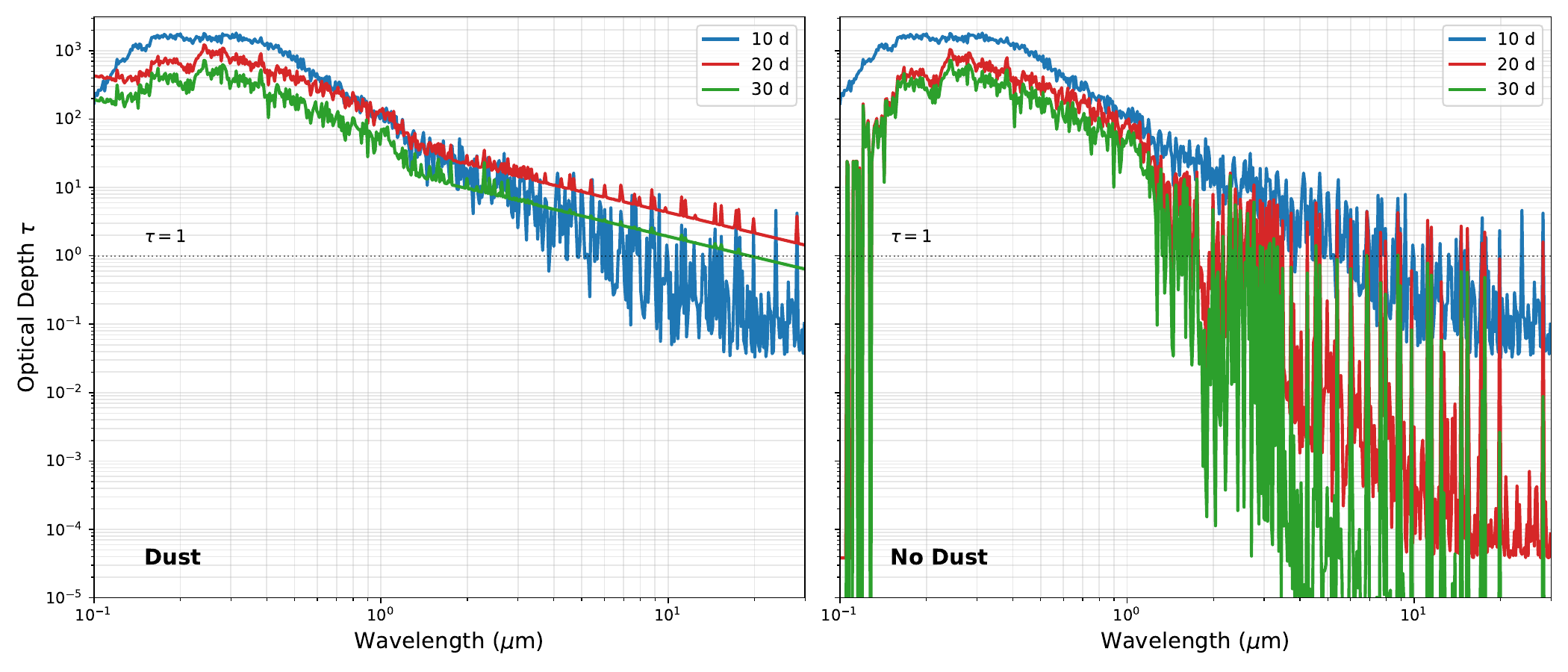}
    \caption{
   Radially optical depth of the ejecta as a function of wavelength for a radiative transfer model including dust (left panel) and not including dust (right panel). When dust is not included, the opacity is provided by blended line transitions, and falls off sharply at longer wavelengths. By 20 days, the dust free ejecta is extremely optically thin in the near infrared ($\sim 5\,\mu$m). When dust is included, the grains provide a smooth continuum opacity which keeps the ejecta optically thick through the near infrared for over 2 months. 
    \label{fig:opacities}
    }
\end{figure*}

\subsection{Spectral Models of Dusty Kilonovae}
The left panel of Figure~\ref{fig:spectrum} shows the evolution of the dust mass fraction as a function of ejecta velocity for a model with a total mass of $\Mej=0.1\,\Msun$ and a kinetic energy of $9 \times 10^{50}$\,erg. By day~20, dust forms efficiently in the denser, inner regions of the ejecta, while the outer layers remain dust-free owing to their low densities.
The dusty interior of the ejecta is optically thick, producing a nearly blackbody continuum that reproduces the observed infrared spectrum of AT2023vfi well, as shown in the right panel of Figure \ref{fig:spectrum}. Our model also shows a smaller emission peak near 2.2\,$\mu$m, similar to that observed in AT2023vfi. This emission arises from the dust-free outer layers of the ejecta, which are slightly hotter than the interior due to inefficient radiative cooling in the optically thin gas, and produce a bluer component of line emission.

Figure~\ref{fig:spectrum_dust_nodust} shows the spectral time series for two radiative transfer calculations based on this ejecta model: one including dust absorption and emission (thick), and one without dust (dashed). The two models produce identical predictions at early times, when the ejecta is too hot for grain formation, but diverge once the temperature drops below the condensation threshold and dust begins to form at $t \gtrsim 10$\,days.

By times $t \gtrsim 20$\,days, the dust-free model departs significantly from a blackbody spectrum. As shown in Figure~\ref{fig:opacities}, the ejecta at these epochs is extremely optically thin ($\tau \lesssim 10^{-4}$) in the NIR continuum. The emission is then dominated by line emission complexes, producing a strongly non-thermal, nebular spectrum, as is typical for transients in this phase. By day~30, there is little flux beyond $3\,\mu$m, and the predicted spectrum bears little resemblance to the JWST observations of AT2023vfi.

We note that the detailed locations of spectral features in the dust-free model may not be reliable. The atomic data used in these radiative transfer calculations are derived from uncalibrated atomic structure models, and the wavelengths and oscillator strengths of the bound-bound transitions are uncertain. In addition, the calculations neglect non-LTE effects, which are expected to be important under optically thin nebular conditions. Nevertheless, the overall non-thermal, line-dominated character of the spectrum resembles that seen in transients during the nebular phase and is consistent with expectations for kilonovae at late times \citep{Hotokezaka2021}.

When dust is included, grain formation begins after $\sim 10$\,days, first in the cooler outer layers of the ejecta (Figure~\ref{fig:spectrum}). By day~20, dust formation has largely saturated throughout the inner regions. The dust mass fraction declines sharply above the transition velocity $v_{\rm t}$ due to the steep density gradient in the outer ejecta. Only a very small fraction of atoms ($f \sim 10^{-7}$--$10^{-8}$) are incorporated into condensates in these layers, indicating that the association process does not proceed significantly beyond the bottleneck of dimer formation. The total mass of dust-poor ejecta (defined  as regions with dust mass fraction $< 10^{-3}$) is $\approx 5.5 \times 10^{-3}\,M_\odot$, corresponding to only 5\% of the total ejecta mass.

In the model including dust, the opacity of grains provides a continuum opacity in the infrared that remains optically thick until $\sim 70$\,days. Emission from dust grains in the interior regions ($v < v_{\rm t}$) produces a nearly thermal blackbody continuum. The dust-free outer layers contribute a weaker emission component, producing a modest peak near $2.2\,\mu$m, similar to a feature observed in AT2023vfi. This feature is weak compared to the blackbody component, as the outer layers contain only a small fraction of the total ejecta mass. The line profile is flat-topped, as expected from a detached shell of emitting material, with the profile wings extending to velocities  (measured relative to 2.2\,$\mu$m) of $v = 0.18\,c$.

The emission peak near $2.2\,\mu$m in kilonova spectra has been associated with forbidden [\ion{Te}{iii}] line emission \citep{Hotokezaka2023}.  The  uncertainties in our transport model (due to incomplete atomic data and the neglect of non-LTE effects) preclude a rigorous line identification. In our calculations, the opacity and emissivity near 2.2\,$\mu$m are due to a complex blend of lines (Figure~\ref{fig:opacities}) presumably from multiple ionic species. In AT2023vfi, the emission profile showed evidence of multiple components \citep[e.g.,][]{Gillanders2025,Pognan2026} indicating a potentially complex geometry or contributions from multiple lines. A detailed analysis and fitting of line features is beyond the scope of this paper; the main takeaway from our calculations is that emission in the dust-free regions of ejecta are likely to produces nebular line superimposed on a blackbody from a dusty photosphere.

By day 60, the continuum emission in the model has moved redwards and the most conspicuous feature in the JWST/NIRSpec range ($\sim 2$--$5\,\mu$m) is the line-like complex near $2.2\,\mu$m produced in the dust-free layers. The  model spectrum is qualitatively similar to the observed spectra of AT2023vfi at this epoch, however the line feature complex is  broader in the model and the blackbody continuum too red. These discrepancies could be presumably be reduced by adjusting the model parameters (e.g., reducing $v_{\rm t}$) but we avoid fine-tuning the model to fit individual events. Future observations of kilonova spectra at wavelengths redder than $5\,\mu$m would be useful for probing the relative contributions of lines and dust.

\section{Discussion}
\label{sec:discussion}

\subsection{Composition Dependence}
Our results show that dust forms efficiently in kilonova ejecta with $X_{\rm ref}\approx3$\%, which can account for the observed emission of AT2023vfi.
Beyond the hydrodynamic parameters of the ejecta, the efficiency of dust formation depends sensitively on the abundance of refractory elements.
As illustrated in Figures~\ref{fig:element} and \ref{fig:abun}, ejecta with $X_{\rm ref}\approx3$\%, consistent with an abundance pattern similar to that observed in the metal-poor $r$-process-enhanced star HD~222925, can form substantial amounts of dust at masses and velocities typical of kilonova ejecta.
By contrast, ejecta with an abundance pattern similar to that of the metal-poor star with a weak $r$-process signature HD~122563 are expected to remain dust-free under similar conditions.
Thus, efficient dust formation of AT2023vfi inferred in this work may provide evidence that the event produced heavy $r$-process elements enriched in third $r$-process peak, HD~222594-like composition as suggested by \citet[Supplementary Figure~10]{Levan2024}.

We note that low-electron-fraction ($Y_{\rm e}$) inner ejecta containing rich third peak elements is not a strict prerequisite for efficient dust formation. Elements such as Zr, Nb, Mo, and Ru ($Z=40$--42, 44), which are synthesized in ejecta with moderate $Y_{\rm e} \sim 0.3$, also have high condensation temperatures (Figure~\ref{fig:element}). Although single-element calculations indicate condensation temperatures near $\sim 1400$\,K, actual grain growth is likely multi-element co-condensation; as illustrated by the alloy case in Figure~\ref{fig:element}, formation of complex elemental mixtures can substantially elevate the effective condensation temperature. Thus, if such light refractory elements dominate the composition, they could theoretically produce a non-negligible amount of dust. While this mechanism remains speculative, the potential for dust formation from lighter refractory elements at moderate $Y_{\rm e}$ offers an open possibility that can be tested by future observations.

\subsection{Diversity in Late-time Kilonovae}
Our results suggest that dust formation is likely to occur $\sim10$--20\,days after the merger. Spectroscopic observations during this phase may capture the transition from spectra dominated by atomic lines to continua dominated by thermal dust radiation (Figure \ref{fig:spectrum_dust_nodust}). Such spectral transitions provides a key diagnostic in kilonova ejecta, analogous to dust-formation signatures observed in nearby Type II supernovae \citep[e.g.,][]{Wooden1993, sn2004et, sn2004dj, Jacobson2025_ixf, Medler2025_ixf, Singh2026_ixf}.

The velocity dependence of dust formation naturally connects to the multiple ejecta components predicted by merger simulations \citep[e.g.,][for a review]{SH2019}.
These include high-velocity dynamical ejecta ($\gtrsim 0.2\,c$), slower post-merger outflows from the accretion disk ($\lesssim 0.1\,c$), and very slow disk-evaporation ejecta at late times \citep[$\sim0.01\,c$; ][]{Lu2023}. Because dust formation is efficient in slower ejecta, the latter two components are likely the primary sites of $r$-process dust production. Variations in ejecta mass, velocity, and composition between mergers should lead to corresponding diversity in late-time kilonova emission, ranging from spectra dominated by atomic lines to continua dominated by thermal dust radiation. 
Future optical and infrared observations will provide a direct test of this scenario, both through time-resolved spectroscopy around the dust-formation epoch and through the diversity of late-time kilonova emission.

Our discussions have been primarily based on the JWST AT2023vfi spectra. However, Spitzer photometry observations of AT2017gfo at 43\,days reported a detection of $\lambda F_\lambda\sim8\times10^{-17}$\,erg\,s$^{-1}$\,cm$^{-2}$ at 4.5\,$\mu$m and an upper limit of $\lambda F_\lambda\sim3\times10^{-17}$\,erg\,s$^{-1}$\,cm$^{-2}$ at 3.6\,$\mu$m when scaled to the distance of AT2023vfi \citep{Villar2018, Kasliwal2022}. In fact, these flux levels are similar to AT2023vfi at 61\,days. The two-points SED fits a blackbody at $\sim500$\,K, which can be reproduced by dust continuum requiring a dust mass roughly an order of magnitude smaller than AT2023vfi at 29\,days \citep[though see, e.g.,][for an interpretation by nebular emission spectrum]{Hotokezaka2022}. Because the late-time dataset for AT2017gfo consists of only two photometric bands at a single epoch, we refrain from detailed modeling. Nevertheless, these comparisons hint at potential diversity of late-time kilonova emission, requiring a larger sample size from future observations to establish the true variation.

\subsection{Effects of Radioactive Particles}
\label{sec:radioactive}
Our calculations have neglected the effects of $\alpha$-particles and fission fragments. However, $\alpha$-decay and fission are potentially important on timescales of $\sim10$--100\,days. Given their large ionization cross sections \citep[e.g.,][]{Hotokezaka2020, vandenBerg2026}, the number of ion pairs and energy deposition by heavy decay products per unit length are much larger than those by $\beta$-particles. As a result,
these heavy particles may critically affect dust formation and survival.

Let us first estimate the impact rates of heavy decay products on grains. Using the number flux density per unit energy $y$, the impact timescale of an $\alpha$-particle or a fission fragment on a spherical dust grain of radius $a$ can be estimated as $t_{\rm imp}\approx (\pi a^2 y\Delta E)^{-1}$. For a representative 5-nm grain and typical energies of heavy particles, we obtain
\begin{align}
	t_{{\rm imp}, \alpha} &\approx 13\,{\rm s}\, 
    \bigg(\frac{y_\alpha}{10^{11}\,{\rm s^{-1}\,cm^{-2}\,MeV^{-1}}}\bigg)^{-1}
    \bigg(\frac{\Delta E}{1\,{\rm MeV}}\bigg)^{-1} \label{eq:alpha}\\
	t_{\rm imp, fs} &\approx 0.35\,{\rm hr}\, 
    \bigg(\frac{y_{\rm fs}}{10^7\,{\rm s^{-1}\,cm^{-2}\,MeV^{-1}}}\bigg)^{-1}
    \bigg(\frac{\Delta E}{100\,{\rm MeV}}\bigg)^{-1},\label{eq:fs}
\end{align}
where we adopt the values of $y_{\alpha}$ and $y_{\rm fs}$ at 10\,days after the merger from \citet{vandenBerg2026}.
These fluxes are proportional to the nuclear heating rates of $\alpha$-decay, $\dot{q}_{\alpha}$, and spontaneous fission, $\dot{q}_{\rm fs}$, respectively. \cite{vandenBerg2026} assume that $\dot{q}_{\alpha}$ is comparable to the $\beta$-decay heating rate from a few days to 100\,days. They further assume that $\dot{q}_{\rm fs}$ is dominated by the spontaneous fission of $^{254}$Cf and becomes comparable to the $\beta$-decay heating rate at 100\,days. Note that these heating rates are sensitive to the $Y_{\rm e}$ distribution and the nuclear models, and can vary by orders of magnitudes \citep[e.g.][]{Wanajo2018,Wu2019}.
These estimates indicate that heavy decay products can impact grains on timescales much shorter than the expansion and grain growth timescales.

\begin{figure}
    \centering
	\includegraphics[width=\linewidth]{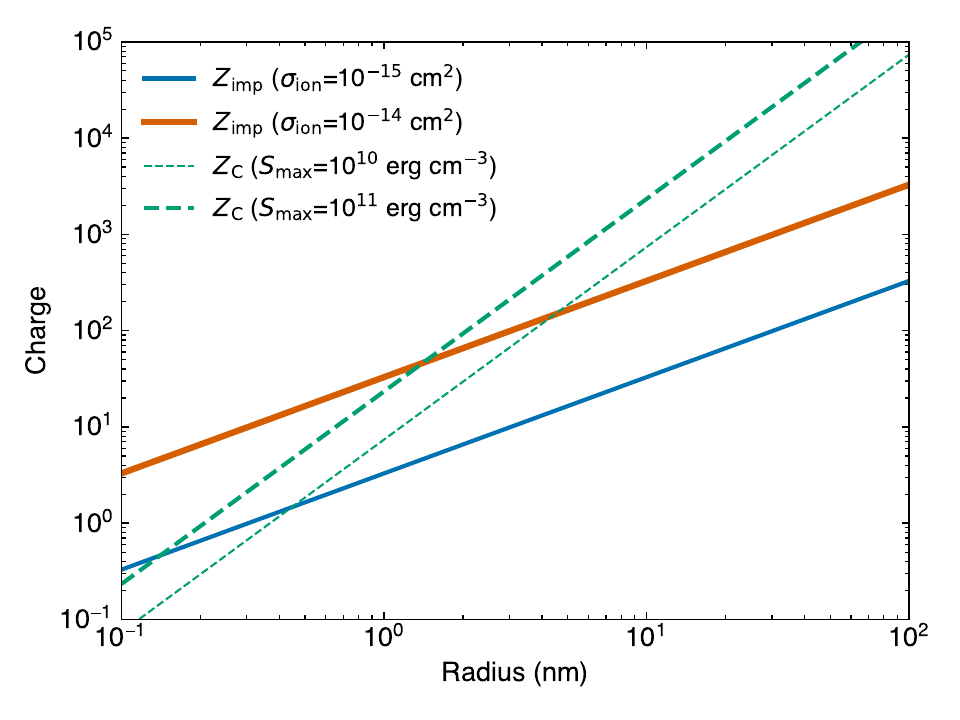}
    \caption{
   Induced charge $Z_{\rm imp}$ on a dust grain as a function of grain radius $a$, compared with the critical disruption charge $Z_{\rm C}$. Solid lines show the charge induced by heavy-particle impacts for ionization cross sections of $\sigma_{\rm ion}=10^{-15}\,{\rm cm}^2$ ($\alpha$-particles; blue) and $10^{-14}\,{\rm cm}^2$ (fission fragments; orange), assuming a grain density of $\rho_{\rm d}=10\,{\rm g\,cm^{-3}}$ and atomic mass number $A=190$. Dashed green lines show the maximum charge $Z_{\rm C}$ against electrostatic disruption for tensile strengths of $S_{\rm max}=10^{10}$ and $10^{11}\,{\rm erg\,cm^{-3}}$. Grains with $Z_{\rm imp}\ge Z_{\rm C}$ undergo disruption.
    \label{fig:Zimp}
    }
\end{figure}

Heavy decay products ionize the surrounding material while undergoing thermalization. Typical ionization cross sections for $\alpha$-particles and fission fragments are $\sigma_{{\rm ion}, \alpha}(E\approx 1\,{\rm MeV})\sim10^{-15}$\,cm$^2$ and $\sigma_{\rm ion, fs}(E\approx 100\,{\rm MeV})\sim10^{-14}$\,cm$^2$, respectively, which are substantially larger than the ionization cross section of $\beta$ particles, $\sigma_{\rm ion, \beta}(E\approx 1\,{\rm MeV})\sim 10^{-18}\,{\rm cm^2}$ \citep{vandenBerg2026}.
Therefore, the number of ion pair created by an impact of heavy decay products on a grain is much larger than that of $\beta$-particles.
Assuming that the ionization events contribute to the net grain charge, the characteristic impact-induced charge $Z_{\rm imp}$ is estimated as
\begin{equation}
    Z_{\rm imp}\approx \sigma_{\rm ion}\frac{\rho_{\rm d}}{Am_u}a.
\end{equation}
For a 5-nm grain with $\rho_{\rm d}=10$\,g\,cm$^{-3}$ and $A=190$, this yields $Z_{\rm imp}\approx 158(\sigma_{\rm ion}/10^{-14}\,{\rm cm}^2)$, shown as solid lines in Figure~\ref{fig:Zimp}.

This induced charge can be compared with the maximum critical charge $Z_{\rm C}$ that a grain can retain without undergoing electrostatic disruption. Using the binding energy per unit volume of grains (so-called tensile strength) $S_{\rm max}$, the critical charge is given by $Z_{\rm C}=a^2\sqrt{4\pi S_{\rm max}}/e$ \citep{Draine1987, Waxman2000}. The dashed lines in Figure~\ref{fig:Zimp} show the calculated criteria assuming representative values of $S_{\rm max}=10^{10}$--10$^{11}$\,erg\,cm$^{-3}$, although the exact values for refractory metals remain unknown. Due to excessive charging, grains in the region of $Z_{\rm imp}\ge Z_{\rm C}$ will experience electrostatic disruption. We find that fission fragments with $\sigma_{\rm ion}\sim10^{-14}$\,cm$^2$ can destroy nanometer-sized grains, while the direct effects of $\alpha$-particles remain marginal. These estimates suggest that dust grains, which are expected to reach typical sizes of a few nanometers in kilonova ejecta, cannot survive under strong fission heating. 

Although the electrostatic disruption caused by $\alpha$-particles is marginal, repeated impacts may still affect grain growth during the early sub-nanometer stage. Comparing the collisional timescale $t_{\rm coll}$ (equation~\ref{eq:tcoll}), evaluated at $T=1800$\,K, with the $\alpha$-particle impact timescale $t_{{\rm imp},\alpha}$, we find that at gas densities above the threshold for efficient dust formation (see Figure~\ref{fig:abun}), the atom-atom collisional timescale can be shorter than or comparable to the $\alpha$-impact timescale for a target cross section of $\pi a^2\sim 10^{-15}$\,cm$^2$, comparable to $\sigma_{{\rm ion}, \alpha}$. When atomic collisions occur much more frequently than $\alpha$-particle impacts, the latter are unlikely to significantly hinder grain growth. However, when the two timescales become comparable, the net effect of repeated impacts on grain growth remains complex and requires further investigation. In particular, these effects depend on the energy deposited in the grain by each impact, which remains uncertain; therefore, the impact frequency alone is insufficient to evaluate their effects on grain growth.

We have also neglected the effects of $\beta$-particles on grains other than dimers. The $\beta$-impact timescale for a 5-nm grain can be estimated as $0.13$\,s, adopting $y_\beta\sim10^{14}$\,s$^{-1}$\,cm$^{-2}$\,MeV$^{-1}$ and $\Delta E\sim0.1$\,MeV at 10\,days \citep[equation~\ref{eq:y}; see also][]{vandenBerg2026}, suggesting that grains may experience numerous impacts during their evolution. As in the case of $\alpha$-particles, however, the effects of these repeated impacts on grain growth and survival depend critically on the energy deposited per impact and the resulting structural damage, which is expected much smaller than that of $\alpha$-particles. A quantitative evaluation of the energy deposition and its effects on grain structure for both $\alpha$- and $\beta$-particles is an important subject for future investigation.

Finally, we emphasize that the above destruction mechanism depends sensitively on the heating rates, which are subject to uncertainties in yields and nuclear models. Lower heating rates than those assumed here would reduce the impact rates and potentially allow grains to grow and survive, whereas higher heating rates could enhance grain disruption.
The presence or absence of late-time thermal dust emission in kilonovae may therefore provide a probe of the contributions of $\alpha$-decay and spontaneous fission to the radioactive heating.
We will explore the details of grain disruption caused by heavy decay products in a future work.

\section{Conclusions}
\label{sec:conclusion}
We have studied the formation and emission of dust composed of $r$-process refractory elements in kilonovae.
By calculating the kinetic evolution of clusters in kilonova ejecta under an empirically derived single power-law temperature evolution, we found that $r$-process refractory elements form dust, particularly in lower-velocity ejecta.
This is in contrast with a previous study that adopted a different approach for modeling dust formation \citep[see Appendix \ref{sec:comparison}]{Takami2014}.
We found that the dusty interior of the ejecta can become optically thick at $\sim30$\,days and produce a nearly blackbody continuum. This plausibly accounts for the observed infrared spectrum of AT2023vfi with $\Mej=0.1\,\Msun$ and $X_{\rm ref}\approx3$\%, \ie with a relatively large fraction of low-$Y_{\rm e}$ material, although further refinement of ejecta parameters and geometry will be required to match all observational features.

We emphasize that our findings are based on the assumption that the ejecta cools sufficiently for dust to form. In addition, reactions involving charged species are not included due to the current lack of relevant kinetic rate data. Determining temperature evolution and ionization structures demands complex non-LTE calculations, which are currently limited by the lack of extensive atomic data for heavy elements. Even if dust grains form successfully, whether they survive depends on the heating rate from heavy particles. Despite these uncertainties, we stress that infrared  emission from $r$-process dust can potentially provide a new diagnostic of heavy-element production in neutron star mergers, testable through time-series spectroscopy and late-time emission diversity.

Finally, similar $r$-process dust formation could also occur in other proposed $r$-process sites such as collapsars \citep[e.g.,][]{Siegel2019, Miller2020, Just2022, Shibata2025}, magnetorotational supernovae \citep[e.g.,][]{Nishimura2017, Mosta2018, Reichert2021}, and accretion-induced collapse \citep{Batziou2025, Pitik2026}. Because these events also produce lighter elements that can form carbonaceous, oxide, or silicate grains, the resulting dust formation and emission may be more complex than that in neutron star mergers. Even so, such dust emission could provide a valuable diagnostic of $r$-process nucleosynthesis in these events, offering an important avenue for future investigation.

\begin{acknowledgments}
We thank J.~Mat\'u\v{s}ka for helping us to use the data of W clusters, and the anonymous referee for critical and constructive comments that improved the paper.
N.D. is supported by the Grants-in-Aid for JSPS Fellows (25KJ0075) and for Scientific Research from JSPS (26K21726). K.H. is supported by the Grants-in-Aid for Scientific Research from JSPS (23H01169, 23H04900, 24KF0202) and the JST FOREST Program (JPMJFR2136). D.K. is supported in part by the U.S. Department of Energy, Office of Science, Office of Nuclear Physics, DE-AC02-05CH11231, DE-SC0004658, and DE-SC0024388, and by a grant from the Simons Foundation (622817DK).
\end{acknowledgments}

\appendix
\section{Abundance Pattern Models of $r$-process Elements}
\label{sec:abun}
\begin{figure*}
	\includegraphics[width=\linewidth]{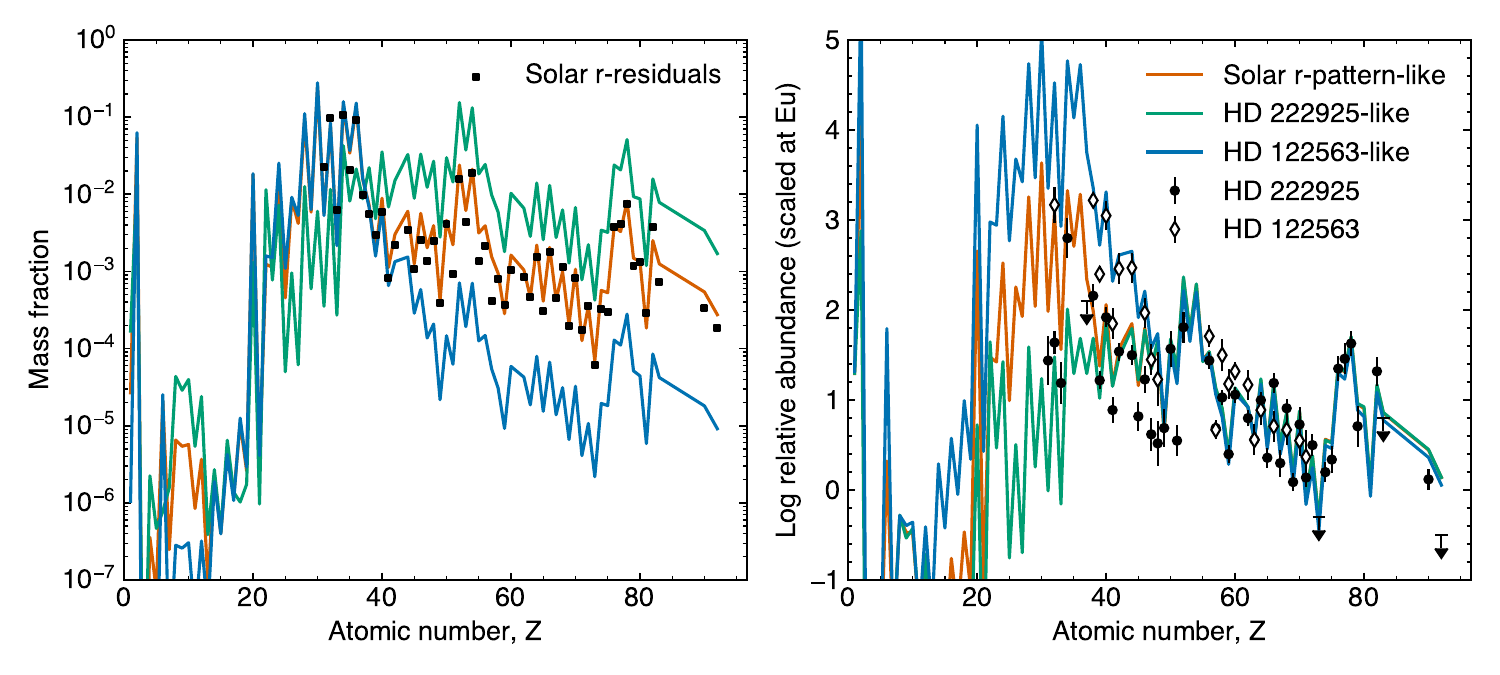}
    \caption{
    Mass fractions (left) and relative abundances (right) as a function of atomic number for different abundance models in kilonova ejecta \citep{Wanajo2018, Domoto2021, Domoto2022}. In the left panel, the solar $r$-residual pattern \citep[squares;][]{Prantzos2020} is scaled to match the solar r-pattern-like model at Eu ($Z=63$). In the right panel, all the abundances are scaled at Eu ($Z=63$).
    }
    \label{fig:elem-model}
\end{figure*}

The right panel of Figure~\ref{fig:element} shows the cumulative mass fractions of abundance models constructed to reproduce the solar $r$-residual pattern \citep{Prantzos2020} and those observed in metal-poor $r$-process-enhanced stars \citep{Roederer2022, Honda2006, Cowan2005, Roederer2012}. The mass fractions and relative abundances of each model as a function of atomic number are shown in Figure \ref{fig:elem-model}. These models are based on a nucleosynthesis calculation of \citet{Wanajo2018} and have been used in \citet{Domoto2021, Domoto2022}. The solar $r$-residual-like model corresponds to the `mFE-a' model of \citet{Wanajo2018}, while the HD~222925-like and HD~122563-like models correspond to the `Solar' and `Light' models of \citet{Domoto2021, Domoto2022}, respectively. Further details of the model construction can be found in the cited references.

\begin{figure*}
	\includegraphics[width=\linewidth]{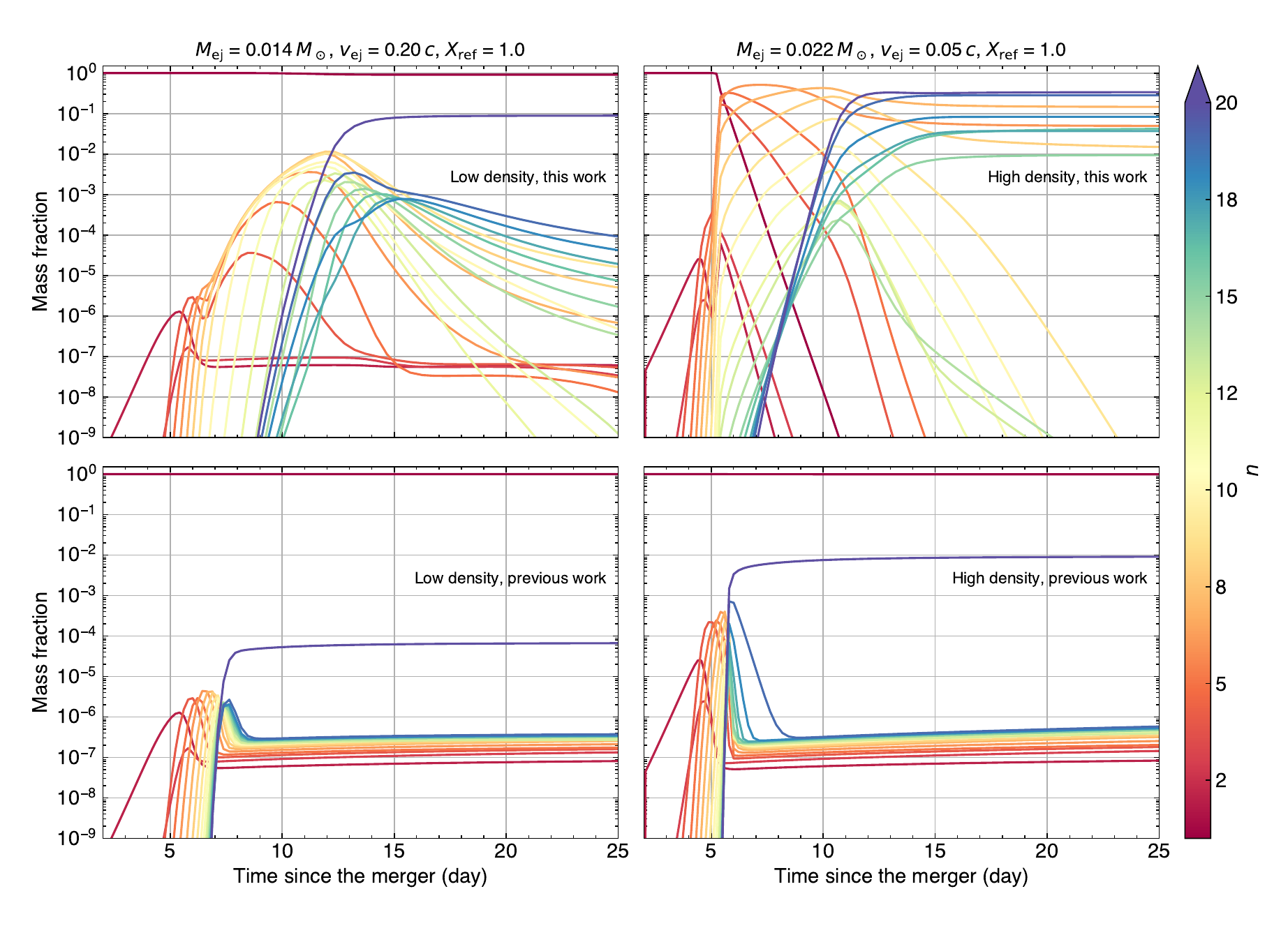}
    \caption{
    Comparison with models configured to mimic those in \citet{Takami2014}, assuming 100\% of the ejecta material is condensable.
    The mass fractions of clusters of different sizes are shown in different colors, as indicated in the color bar.
    The left and right panels show the low- and high-density cases, respectively, with $\Mej$ and $\vej$ indicated above each panel.
    The top panels show the results including all reactions, while the bottom panels show those including only monomer reactions, configured to mimic the approach of \citet{Takami2014}.
    }
    \label{fig:takami}
\end{figure*}

\section{Comparison with previous work}
\label{sec:comparison}
\citet{Takami2014} studied dust formation in kilonovae, and concluded that dust grains of $r$-process elements are difficult to form.
To compare with their results, we solve the kinetic rate equation for models configured to mimic their low- and high-density cases.
Although they distinguish these cases by treating the opacity as a parameter, we instead set the ejecta density as a function of $\Mej$ and $\vej$ to match the density at 7\,days used in \citet{Takami2014}.
The temperature evolution is also taken to be the same as in that work.
Here we ignore the effects of $\beta$-electrons, and adopt $\kdi =10^{-17}$\,cm$^3$\,s$^{-1}$.
\citet{Takami2014} showed that, in the low-density case, dust grains of $r$-process elements (with Pt used as a representative) do not form even if 100\% of Pt is present in the ejecta.
In contrast, our calculations show the grain formation with a condensation mass fraction of $\sim10$\% in the low-density case, and nearly 100\% in the high-density case, in contradiction to their conclusions (top panels of Figure~\ref{fig:takami}).

The main difference between our calculations and those of \citet{Takami2014} is that we consider association and fragmentation for clusters of all sizes, \ie there are many routes for clusters to grow.
In \citet{Takami2014}, grains grow by adding monomers and erode by removing monomers, \ie only the reaction W + W$_{n-1}$ $\leftrightarrow$ W$_n$ is considered.
To understand the impact of this treatment, the bottom panels of Figure~\ref{fig:takami} show the results including only monomer reactions.
The resulting condensation mass fractions are $\approx 7\times10^{-5}$ and $\approx 0.09$ in the low- and high-density cases, respectively, broadly consistent with those presented in \citet{Takami2014}.
This demonstrates the importance of including all possible reactions for large clusters, rather than only monomer reactions.

\begin{figure*}
	\centering
	\includegraphics[width=0.6\linewidth]{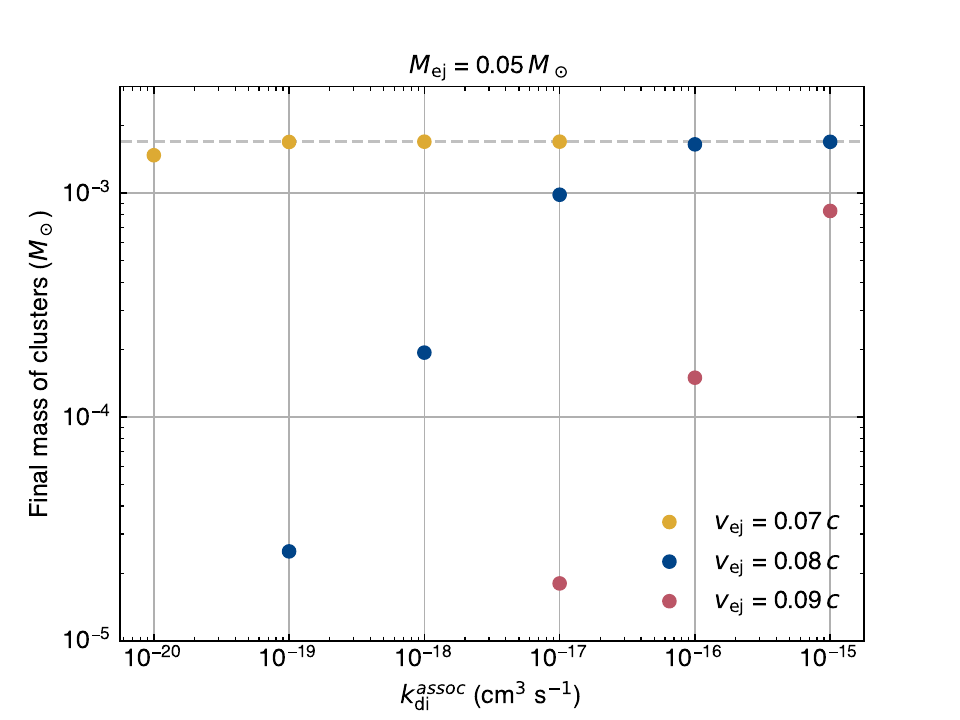}
    \caption{
    Final mass of clusters as a function of $\kdi$ for uniform-density ejecta with a fixed $\Mej=0.05\,\Msun$ and $\vej=0.07\,c$, 0.08\,$c$, and 0.09\,$c$. The horizontal dashed line shows the saturation mass limit, corresponding to the total mass of condensable refractory elements, $\approx 3$\% of $\Mej$.
    }
    \label{fig:k_di}
\end{figure*}

\section{Dependence of dust mass on dimer formation rate}
\label{sec:k_di}
Figure~\ref{fig:k_di} shows the final mass of clusters as a function of $\kdi$ for a fixed $\Mej=0.05\,\Msun$ and selected $\vej$, calculated as in Section \ref{sec:form}. For $\vej\ge0.08\,c$, where $t_a$ tends to be longer due to low densities, the final dust mass scales almost linearly with $\kdi$. On the other hand, for $\vej\le0.07\,c$, the final dust mass depends only weakly on $\kdi$ due to higher densities that remain above the threshold of efficient dust formation. This indicates that while the precise boundary separating dust-rich and dust-free regimes shifts slightly with $\kdi$, the overall trend in dust formation efficiency shown in the right panel of Figure~\ref{fig:abun} robustly holds.

\bibliographystyle{aasjournalv7}
\bibliography{references}

\end{document}